# Random Sampling High Dimensional Model Representation Gaussian Process Regression (RS-HDMR-GPR) for representing multidimensional functions with machine-learned lower-dimensional terms allowing insight with a general method

Owen Ren,[a,b] Mohamed Ali Boussaidi,[a,c] Dmitry Voytsekhovsky,[b] Manabu Ihara[d] and Sergei Manzhos[d,1]

[a] Centre Énergie Matériaux Télécommunications, Institut National de la Recherche Scientifique, 1650 boulevard Lionel-Boulet, Varennes QC J3X1S2 Canada.

[b] PureFacts Inc., 48 Yonge St Suite 400, Toronto, ON M5E 1G6 Canada.

[c] Ecole Nationale d'Ingénieurs de Tunis, Rue Béchir Salem Belkhiria Campus universitaire, BP 37, 1002, Le Bélvédère, Tunis, Tunisia.

[d] School of Materials and Chemical Technology, Tokyo Institute of Technology, Ookayama 2-12-1, Meguro-ku, Tokyo 152-8552 Japan

## Abstract

We present an implementation for the RS-HDMR-GPR (Random Sampling High Dimensional Model Representation Gaussian Process Regression) method. The method builds representations of multivariate functions with lower-dimensional terms, either as an expansion over orders of coupling or using terms of only a given dimensionality. This facilitates, in particular, recovering functional dependence from sparse data. The method also allows for imputation of missing values of the variables and for a significant pruning of the useful number of HDMR terms. It can also be used for estimating relative importance of different combinations of input variables, thereby adding an element of insight to a general machine learning method, in a way that can be viewed as extending the automatic relevance determination approach. The capabilities of the method and of the associated Python software tool are demonstrated on test cases involving synthetic analytic functions, the potential energy surface of the water molecule, kinetic energy densities of materials (crystalline magnesium, aluminum, and silicon), and financial market data.

[1] Corresponding author. E-mail: manzhos.s.aa@m.titech.ac.jp

**Keywords**: multivariate function, machine learning, Gaussian process regression, High Dimensional Model Representation, data imputation.

## 1 Introduction

Interpolation or fitting of a multivariate function $f(\boldsymbol{x}), \boldsymbol{x} \in R^D$ from discrete samples $\{\boldsymbol{x}^{(j)}\}, j = 1,2,\dots,M$ (forming a matrix of features $\boldsymbol{X}$ with $M$ rows and $D$ columns and a column vector of $M$ targets $\boldsymbol{Y}$) is a ubiquitous problem in natural sciences and beyond. In high-dimensional spaces (with dimensionality $D$ of more than six), this problem is difficult enough so that standard approaches such as polynomial interpolation, splines or direct product basis expansions (e.g. Fourier expansion) often fail. The difficulties of applying such standard methods in high-dimensional spaces have to do, in particular, with an exponential growth of the required numbers of sampled values and of terms in direct product type representations. These numbers quickly become impractical as $D$ is increased; this is known as the "curse of dimensionality".[1] The sampling of functions in high-dimensional spaces is necessarily sparse; for example, 100,000 samples in a 20-dimensional space is equivalent to about 1.8 points per dimension of a direct product grid. Increasing the number of data points to 1,000,000 would only result in a density of sampling increase to about 2 points per dimension. Methods that can work well with sparse samples are therefore desired. Furthermore, the distribution of sampled data may be arbitrary, and data, especially those collected during observation of physical, social, and other types of phenomena, do not generally lie on a grid or on sub-dimensional hypersurfaces and do not follow any particular distribution. Methods that can work well with any distributions of data are therefore desired. It is clear from the above that in high-dimensional spaces, non-direct product approaches need to be used. Machine learning approaches such as neural networks (NN)[2, 3] or Gaussian process regression (GPR)[4] used in the present work are examples of such representations which in this sense avoid the exponential scaling of the cost (both in terms of required data and in terms of required number of terms in the representation) with dimension.

Even with machine learning methods, the problem of reconstructing functional dependence from sparse, arbitrarily distributed samples in multidimensional spaces is difficult. For example, the number of required data and neurons of an NN may be large and lead to overfitting even as NNs can handle millions of data,[5, 6] while the cost of GPR becomes problematic with more than about 10,000 data points (due to the need to operate with a matrix of size $M \times M$ and respective matrix-vector products). A useful approach that leverages the capability of ML methods is representation of a multidimensional function with lower-dimensional terms. Rabitz et al. introduced High Dimensional Model representation (HDMR)[7-9] which is an expansion over orders of coupling:

$$f(\boldsymbol{x}) \approx f_0 + \sum_{i=1}^{D} f_i(x_i) + \sum_{1 \le i < j \le D} f_{ij}(x_i, x_j) + \cdots + \sum_{\{i_1 i_2 \dots i_d\} \in \{12 \dots D\}} f_{i_1 i_2 \dots i_d}(x_{i_1}, x_{i_2}, \dots, x_{i_d}) \tag{1}$$

where $d \le D$ is the maximum order of coupling considered. The ANOVA decomposition[10, 11] used in data analysis and the many-body and n-mode approximations[12-16] used in computational chemistry are particular cases of HDMR. In many practically important applications, using terms with $d \le 3$ is often sufficient.[7] Moreover, when sampling is sparse, only low-dimensional component functions can be reliably recovered.[17, 18] Fitting/interpolation of a lower-dimensional component function $f_{i_1 i_2 \dots i_d}(x_{i_1}, x_{i_2}, \dots, x_{i_d}), d < D$, in particular with ML methods, is in principle an easier task than constructing the full-dimensional $f(\boldsymbol{x})$. A representation with lower-dimensional functions is also advantageous when using the function in applications; this advantage is particularly pronounced if $f(\boldsymbol{x})$ needs to be integrated.

One advantage of HDMR as introduced by Rabitz et al. (as well as of the ANOVA and n-mode representations) is orthogonality of its component functions $f_{i_1 i_2 \dots i_d}(x_{i_1}, x_{i_2}, \dots, x_{i_d})$:

$$\int_D w(\boldsymbol{x}) f_{i_1 i_2 \dots i_d}(x_{i_1}, x_{i_2}, \dots, x_{i_d}) f_{j_1 j_2 \dots j_m}(x_{j_1}, x_{j_2}, \dots, x_{j_m}) d\boldsymbol{x} = 0$$
$$\{i_1, i_2, \dots, i_d\} \ne \{j_1, j_2, \dots, j_m\} \tag{2}$$

when the weight function $w(\boldsymbol{x})$ is of the form $w(\boldsymbol{x}) = \prod_{i=1}^{D} w_i(\boldsymbol{x_i})$, i.e. when it is a product of individual weight functions of a single variable. When the samples are constrained to lie on *d*-dimensional hypersurfaces spanned by each variable combination $\left(x_{j_1}, x_{j_2}, \dots, x_{j_m}\right)$, one obtains the so-called cut-HDMR which in chemical physics has been known as the n-mode expansion.[7, 9] An advantage of cut-HDMR is that the component functions are then easily defined as "slices" of $f(\boldsymbol{x})$ obtained by fixing some of the coordinates to the co-called expansion center. An obvious disadvantage of such an approach is the need for a separate dataset for each component function, and the number of those scales combinatorically with *d* and *D*. If the data are distributed (e.g. randomly) in the *D*-dimensional space (as opposed to being constrained to *d*-dimensional hypersurfaces[7, 9]), the component functions of the resulting Radom Sampling (RS-) HDMR[19, 20] can all be determined from one and the same dataset and are defined via multidimensional integrals, specifically, $(D - d)$ dimensional for *d*-dimensional component functions. (We use "RS" following the definition of Rabitz et al. The data distribution need not be random; rather, the method does not constrain the points to be distributed in any particular way, as opposed to the cut-HMDR / n-mode scheme where the points must lie on sub-dimensional hypersurfaces.) The integrals approach is obviously impractical for high *D*.[8, 19-21] We previously introduced a generalization of (RS) HDMR where $f(\boldsymbol{x})$ is approximated as[17, 18, 22]

$$f(\boldsymbol{x}) \approx \sum_{\{i_1 i_2 \dots i_d\} \in \{12 \dots D\}} f_{i_1 i_2 \dots i_d}(x_{i_1}, x_{i_2}, \dots, x_{i_d}) \tag{3}$$

i.e. as a sum of lower-dimensional functions of any given dimensionality $d < D$. The component functions $f_{i_1 i_2 \dots i_d}(x_{i_1}, x_{i_2}, \dots, x_{i_d})$ of this representation can be constructed by fitting sequentially each component function in cycles as

$$f_{k_1 k_2 \dots k_d}\left(x_{k_1}, x_{k_2}, \dots, x_{k_d}\right) = f(\boldsymbol{x}) - \sum_{\substack{\{i_1 i_2 \dots i_d\} \in \{12 \dots D\} \\ \{i_1 i_2 \dots i_d\} \neq \{k_1 k_2 \dots k_d\}}} a(c) f_{i_1 i_2 \dots i_d}(x_{i_1}, x_{i_2}, \dots, x_{i_d}) \tag{4}$$

with a generic algorithm like NN or GPR.[17, 18, 22] The factor $a(c)$, where $c$ is the cycle number, is introduced to prevent "crowding out" during the first cycles of other component functions by the component functions which are fitted first (which may happen, in particular, because of the non-unique way in which lower order terms of Eq. 1 can be subsumed into $d$-dimensional terms of Eq. 3), which may lead to the algorithm getting stuck at a local minimum.[18] One starts with a choice of $0 < a(0) < 1$ and brings $a(c)$ to 1 over a number of cycles. This approach does not ensure orthogonality of component functions but gains in simplicity. Specifically, one completely dispenses with the need to compute multi-dimensional integrals, and any distribution of data in the $D$-dimensional space can be used. Building a full expansion over orders of coupling (e.g. if one desires access separately to $f_i(x_i)$ and $f_{ij}(x_i, x_j)$ when $d \geq 3$) is also possible based on Eq. 3, in which case one could fit

$$f_{k_1 k_2 \dots k_d}\left(x_{k_1}, x_{k_2}, \dots, x_{k_d}\right) = T(\boldsymbol{x}) - \sum_{\substack{\{i_1 i_2 \dots i_d\} \in \{12 \dots D\} \\ \{i_1 i_2 \dots i_d\} \neq \{k_1 k_2 \dots k_d\}}} a(c) f_{i_1 i_2 \dots i_d}(x_{i_1}, x_{i_2}, \dots, x_{i_d}) \tag{5}$$

where

$$T(\boldsymbol{x}) = f(\boldsymbol{x}) - \left( f_0 + \sum_{i=1} f_i(x_i) + \sum_{1 \leq i < j \leq D} f_{ij}(x_i, x_j) + \cdots + \sum_{\{i_1 i_2 \dots i_{d-1}\} \in \{12 \dots D\}} f_{i_1 i_2 \dots i_{d-1}}\left(x_{i_1}, x_{i_2}, \dots, x_{i_{d-1}}\right) \right) \tag{6}$$

Alternatively, one can fit terms of different dimensionality simultaneously, which is also allowed in the implementation we use, see section 2.1.

The classic cut-HDMR expansion has recently been combined with GPR in specific applications.[23, 24] The method and code presented here implements Eqs. 1, 4-6 where the component functions $f_{i_1 i_2 \dots i_d}(x_{i_1}, x_{i_2}, \dots, x_{i_d})$ are represented with Gaussian process regression models and is therefore called RS-HDMR-GPR. The reader is referred to the literature[4] for a description of the GPR method which will not be repeated here. A brief summary is given in the Supplementary Material file. GPR[4] is a powerful technique that has been shown to outperform

NNs in accuracy and/or required number of data to achieve a given accuracy. In fact, it is the only ML regression technique that has been shown to do so in *controlled* comparisons.[25, 26] We have recently shown that the RS-HDMR-GPR combination can provide highly accurate fits to sparse multidimensional data when fitting molecular potential energy surfaces used, in particular, to compute highly accurate vibrational spectra, which is a very stringent test.[18] In Ref. [18], we introduced the idea of RS-HDMR-GPR and used it to test a particular case of Eq. 3. Here, we use an arbitrary HDMR-GPR expansion (either as a full expansion of Eq. 1 or combining any component functions of any dimensionality in a desired way) coded in an end-user library with a GUI and demonstrate additional capabilities of the RS-HDMR-GPR approach, including pruning of HDMR terms and combining at will component functions of different dimensionality (including using linear combinations of the original variables) as well as imputation of missing data values, for diverse applications.

The problem of missing data plagues data analysis. We define here missing data as elements of $\boldsymbol{X}$ for which the values are unknown. We assume that all elements of $\boldsymbol{Y}$ are known. The origins of missing data are multiple and ubiquitous. For example, in financial markets data considered below, different trading days for different ticker symbols (such as stocks or indices from different jurisdictions) result in some incomplete rows of $\boldsymbol{X}$ (where each row corresponds to a trading day). In clinical data, as another example, one often aggregates data from many hospitals (as any given hospital often does not have a statistically significant record on a given disease or drug) where some values (patient / test data or various symptoms or reactions to a drug) are recorded and some are not depending on the hospital or on the jurisdiction. The resulting combined data set can have a large number of missing data.

In some cases, the missing values can be handled relatively painlessly. For example, in financial markets data, even when retaining only complete data rows, most of the data would be preserved. Interpolating values on trading holidays is also reasonable. In other cases, including the clinical data example above, this cannot be done. Retaining only the complete rows poses the risk of losing most of the data even if only a small fraction of entries in matrix $\boldsymbol{X}$ are missing. This will happen when many rows have at least one missing value. For example, if one collects 10,000 50-dimensional instances for a total of 500,000 entries of $\boldsymbol{X}$, when as few as 1% of entries are missing, retaining only complete rows would result in discarding up to half of the data if the missing values are spread across the rows and columns evenly (if the missing values are concentrated in certain columns, one can instead discard the variables); 2% of missing entries could make discard all of the data! This is a major issue with relatively little offered by way of remedies. The popular approach of imputing missing values based on data distribution[27] is masking rather than solving the problem. We will show below that the RS-HDMR-GPR approach allows for reliable imputation of missing values for a specific case of one missing value per row of $\boldsymbol{X}$ when the quality of the uncoupled approximation ($d$ = 1) is reasonable, and the component functions are well-behaved.

The RS-HDMR-GPR algorithm and its implementation, including imputation of missing data, are presented in section 2. In section 3, we show examples of its application. The first example is fitting of a molecular potential energy surface (PES). PES construction is important in molecular physics and the availability of a PES permits (and is often indispensable for) studies of phenomena that can be described by quantum or classical dynamics of the nuclei (such as vibrational spectra and reactions).[28] The second example is the fit of kinetic energy density (KED) of real materials from electron density-dependent descriptors. The ability to predict KED from electron density is important for the construction of kinetic energy functionals that are needed for large-scale ab initio modeling. We show that HDMR-GPR can be used, in particular, to identify important terms of the HDMR expansion (i.e. important combinations of variables) and to significantly reduce the number of required HDMR terms.[29] In this way, it presents more extensive capabilities than the automatic relevance determination (ARD) approach that can be achieved the with a plain GPR. This example demonstrates that HDMR-GPR, while being a general method, can generate insight. This is an important quality as machine learning methods are often used for their black-box nature and ability to learn from implicit information contained in the data, which is often done at the price of lack of insight (as opposed to e.g. fitting with physically motivated functions). The third example illustrates HDMR-GPR method's missing value imputation capabilities on samples of different analytic functions. Finally, the fourth example shows the use of HDMR for predicting financial market data and also uses imputation.

## 2 The algorithm

### *2.1 RS-HDMR-GPR algorithm*

The algorithm works as follows: given a feature matrix $\boldsymbol{X}$ of dimension $M \times D$, we specify $N$ matrices $A_1, A_2, \ldots, A_N$ each with $D$ rows and $d$ columns, where $N$ is the number of the component functions (e.g. $N = C_d^D$ is the number of all $d$-dimensional component functions, where $C_d^D$ is the binomial coefficient). These matrices are used to select the inputs for the HDMR component functions by right multiplication with $\boldsymbol{X}$. For instance, if we wanted the component function inputs for $d = 1$ (the one-dimensional HDMR) for a $M \times 3$ feature matrix $\boldsymbol{X}$, then we define

$$\boldsymbol{A}_1 = \begin{pmatrix} 1 \\ 0 \\ 0 \end{pmatrix}, \quad \boldsymbol{A}_2 = \begin{pmatrix} 0 \\ 1 \\ 0 \end{pmatrix}, \quad \boldsymbol{A}_3 = \begin{pmatrix} 0 \\ 0 \\ 1 \end{pmatrix}, \tag{7}$$

and the resulting matrix product $\boldsymbol{X}\boldsymbol{A}_i$ will yield the input for the $i$-th HDMR component function for $i = 1, 2, 3$. The inputs are produced similarly for $d > 1$, in which case we would need more than one column and 1's entered to the matrices $\mathbf{A}_n$, as shown in Eq. 8 below. Note that this way of selecting inputs is not just limited to selecting columns of $\boldsymbol{X}$. By filling the matrix $\boldsymbol{A}$ with other values, we may select any linear combinations of columns of $\boldsymbol{X}$ for each components' input; an example of this is also given in section 3.1 (Eq. 17). Also possible with this type of fitting is the mixing of components of different dimension. For example, for the same above example of a $M \times 3$ feature matrix $\boldsymbol{X}$, we can define

$$\boldsymbol{A}_1 = \begin{pmatrix} 1 \\ 0 \\ 0 \end{pmatrix}, \boldsymbol{A}_2 = \begin{pmatrix} 0 \\ 1 \\ 0 \end{pmatrix}, \boldsymbol{A}_3 = \begin{pmatrix} 0 \\ 0 \\ 1 \end{pmatrix}, \boldsymbol{A}_4 = \begin{pmatrix} 1 & 0 \\ 0 & 1 \\ 0 & 0 \end{pmatrix}, \boldsymbol{A}_5 = \begin{pmatrix} 1 & 0 \\ 0 & 0 \\ 0 & 1 \end{pmatrix}, \boldsymbol{A}_6 = \begin{pmatrix} 0 & 0 \\ 1 & 0 \\ 0 & 1 \end{pmatrix} \tag{8}$$

which will result in fitting all component functions of the approximation $f(\boldsymbol{x}) = \sum_{i=1}^{D} f_i(x_i) + \sum_{1 \le i < j \le D} f_{ij}(x_i, x_j)$ simultaneously (as opposed to fitting sequentially, first first-order terns and then second-order terms). The user may also decide to use only subsets of component function of a given dimension. The choice of which matrices are used is controlled by the user with the only restriction that each matrix must have number of rows equal to $M$ and number of columns less than or equal to $D$.

With the inputs of the component functions defined, the individual GPR models representing the component functions are trained one at a time for $i = 1, 2, \ldots, N$ and in this specified order (the order itself is not important) similarly to Eq, 4:

$$f^{GPR}_{k_1k_2\ldots k_d}\left(x_{k_1}, x_{k_2}, \ldots, x_{k_d}\right) = f(\boldsymbol{x}) - \sum_{\substack{\{i_1 i_2 \ldots i_d\} \in \{12\ldots D\} \\ \{i_1 i_2 \ldots i_d\} \neq \{k_1 k_2 \ldots k_d\}}} a(c) f^{GPR}_{i_1 i_2 \ldots i_d}(x_{i_1}, x_{i_2}, \ldots, x_{i_d}) \tag{9}$$

Each GPR fit invokes the GPR implementation from the Python scikit-learn library (sklearn.gaussian_process.GaussianProcessRegressor[30]) for $d$ dimensions, with the same options available for individual GPR fits with respect to the choices of kernels, noise variance, etc. (see the Manual of the attached code for instructions on invocation). This training is repeated for a number of cycles (which by default is set to 50) until the output of each component function has converged. A priori, it is not known how many cycles it would take for convergence, and this part must be determined empirically by trying various number of cycles. Usually when $N$ is big, more cycles are needed. To prevent crowding out of component functions as described above, the first several cycles use only a percentage of the output as the fit target and slowly across cycles, that percentage is increased to 100. The increase happens linearly until this percentage reaches 100. The rate of increase is controlled by parameters from the *train* function (see the manual). Note that we did not necessarily have to define the increase linearly, any function that monotonically increases reasonably would have worked. We picked a linear function for simplicity. Denote by $a(c)$ this linear growth function as a function of the current cycle $c$. Initially, we set the outputs (labels) of all the component function to equal the vector $\boldsymbol{Y}/_N$. Following that, we fit each component function $f_{i_1 i_2 \ldots i_d}$ for all $\{i_1 i_2 \ldots i_d\} \in \{12 \ldots D\}$ one at a time as per Eq. 9. When a $f_{i_1 i_2 \ldots i_d}$ is fitted, we replace the output of $f_{i_1 i_2 \ldots i_d}$ by $a(c) f_{i_1 i_2 \ldots i_d}$ and continue with the next fit. In the code, we defined $a(c)$ by:

$$a(c) \coloneqq \max\left\{s + \frac{(1-s)ec}{C}, 1\right\} \tag{10}$$

where $s, e$ are specified parameters, $c$ is the current fitting cycle and $C$ is the total number of cycles. The final output of the model is returned as the sum of these component functions. There is also the option of returning the standard deviation of the estimate of $f(\boldsymbol{x})$ computed as a square root of the sum of the variances of the component functions, where for the GPR of each component function the variance of the estimate is given by

$$\Delta f_{i_1 i_2 \ldots i_d}(x_{i_1}, x_{i_2}, \ldots, x_{i_d}) = K^{**} - \boldsymbol{K}^{*}\boldsymbol{K}^{-1}\boldsymbol{K}^{*T} \tag{11}$$

where the matrices ***K*, *K***** and *K*** are computed from pairwise covariances among the data (see Supplementary Material and Refs. 4). Note that Eq. 11 computes the *confidence* of the expectation value of the component function computed by GPR. It is not necessarily a number indicative of the quality of the fit. For example, the confidence interval estimated with Eq. 11 will in general be higher in a full-dimensional fit than with HDMR with a low *d*, even if the RMSE of the fit is higher with a low-*d* model. This is natural, as the values of the first-order HMDR model's component functions are more reliably estimated even with fewer data, even if the 1*d*-HDMR-GPR model overall has a higher fit RMSE due to the neglect of coupling. Values returned by Eq. 11 should therefore not be used as error bars on the fitted values. This is true in any GPR fit; for example, adding Gaussian noise with zero mean to the data will not, as long as data are abundant, make less reliable the estimates of *the expectation values* of $f(\boldsymbol{x})$ even though such values will be much different from the data due to the noise, resulting in a higher RMSE.

## *2.2 Handling of missing data*

The HDMR structure allows for imputation of missing data. We consider a specific case where a single missing value may occur in a row (in any number of rows). This will be an easier to handle case with HDMR but, as shown in the example given in the Introduction, for a given fraction of missing entries, this seemingly simple case represents the worst-case scenario if one has to retain only complete rows. Consider a first order HDMR approximation

$$f(\boldsymbol{x}) \approx f_0 + \sum_{i=1}^{D} f_i(x_i). \tag{12}$$

As shown previously,[18, 31] low-order component functions require fewer data to be well-defined. Specifically, first-order component functions require the fewest data to be reliably constructed. They can be constructed from a subset of *M* data rows, for example from a subset constructed by discarding rows with missing values. After the first-order approximation of Eq. 12 has been built, in particular with a method which allows evaluation of component functions $f_i(x_i)$ at any $x_i$ such as RS-HDMR-GPR in the present work (or the previously introduced RS-HDMR-NN[22, 31] where the component functions are fitted with NNs, although other methods can be used as well),[18] one can evaluate $f_i(x_i^{(m)})$ for any row $m$ where the component $x_i$ is not known:

$$f_i\left(x_i^{(m)}\right) = f\left(\boldsymbol{x}^{(m)}\right) - f_0 - \sum_{\substack{j=1,\\ j\neq i}}^{D} f_j(x_j^{(m)})$$

(13)

from where one imputes

$$x_i^{(m)} = f_i^{-1}(y_i^{(m)})$$

(14)

where $y_i^{(m)} = f_i\left(x_i^{(m)}\right)$ is the argument of the inverse function which is now known (from Eq. 13). If the inverse function $f_i^{-1}$ is single-valued, i.e. if $f_i$ is monotonic, this imputation is as good as the first order HDMR approximation. If the first order HDMR approximation is very accurate, the imputation can be arbitrarily accurate; in general, it is approximate but can be much more precise than a distribution-based imputation. The component functions $f_i$ are in general not monotonic; in this case, Eq. 14 results in several alternative choices for $x_i^m$ and an algorithm is required to pick a choice. The pick may be facilitated if some choices returned by Eq. 14 are deemed to be unrealistic (based on application-specific knowledge) as well as by the use of variable distributions. This will be illustrated in the example below.

If in a given row more than one coordinate value is missing, Eqs. 13, 14 are by themselves insufficient. They could still be used if some missing values are first imputed by other means so that a single missing value per row remained and could be recovered with Eqs. 13, 14. Alternatively, the approach could potentially be extended to a second order HDMR for pairs of missing variables per row, etc. These more complex cases are not treated here. Instead, we will show below that Eqs. 13, 14 already provide significant power of imputation of missing entries.

In our implementation, we represent the inverses $f_i^{-1}$ as a dictionary by subdividing the interval [0,1] into a specified number $s$ of evenly spaced subintervals $[I_i,\ I_{i+1}]$, where $I_i = \frac{i}{s}$ for $i = 0,1, \dots, s-1$ . The dictionary has inputs $I_i$ for all $i$, and the output of each $I_i$ is a tuple $(f_1(\ I_i)$, $f_2(I_i), \ldots, f_s(I_i))$. In the code, this is stored in a lookup table. Given an output value say $y_1$, we select amongst all $i$ the closest $f_1(I_i)$ to $y_1$ and take $I_i$ as the corresponding input. If we have several $f_1(I_i)$ values that are all very close to $y_1$, say within some threshold distance δ, then all of them would be included as candidates. This threshold value is controlled for in the code. If the original functions $f_i$'s are not one-to-one, then we are faced with an issue of deciding which input to pick. In fact, there is no way of knowing a priori which input value is correct, as the best choices are application-dependent. By default, we always pick the first candidate in our implementation.

The code implementing the RS-HDMR-GPR algorithm is provided in the Supplementary Material and the latest version can also be downloaded from https://github.com/owen-ren0003/rshdmrgpr .

# 3 Examples

We present in this section three examples to demonstrate the use of the RS-HDMR-GPR code and the 1*d*-HDMR based missing data handler. The first subsection contains an example from chemistry and is the potential energy surface (PES) of $H_2O$. The second subsection shows that RS-HDMR-GPR can be used to get insight into relative importance of different combinations of input variables on the example of fitting of kinetic energy densities. The third subsection deals with a set of synthetic datasets to demonstrate the use of the 1*d*-HDMR imputation. Finally, we conclude with a relatively high-dimensional example from quantitative finance.

## *3.1 Reconstruction of the potential energy surface of $H_2O$*

In this subsection, we demonstrate the application of the RS-HDMR-GPR algorithm to data encountered in computational chemistry. The example we choose is fitting of a potential energy surface, in this example of the water molecule. We refit an analytic PES of Ref. [32]. Our goal here is not to analyze the PES or to compare the accuracy PES fitting methods, but instead to demonstrate how the RS-HDMR-GPR approach with different orders $d$ and different matrices $\boldsymbol{A}$ can be used to fit such functions and to analyze the trend in fit quality with the order of HDMR and the shapes of component functions. A higher-dimensional example of the PES (for formaldehyde – 6 dimensions and $UF_6$ - 15 dimensions) was given in Ref. [18] including vibrational spectrum calculations which show that the HDMR-GPR method can produce highly accurate PES fits.

The dataset sampling the PES of the water molecule consists of 10,001 rows (data points) and 4 columns. The first three columns are the features (in this the two OH bond lengths and the HOH angle), and the last column contains the target, in this case the value of the potential energy (ranging from 0 to 20,000 $cm^{-1}$). The reader is referred to Refs. [6, 22] for the description of the data generation procedures. The data are included in the Supplementary Material (with the code). The features and the target were linearly scaled to [0,1]. This preserves the distribution of points and non-linear relationships only up to an affine transformation. We train three different models (1*d*-HDMR-GPR, 2*d*-HDMR-GPR as per Eq. 3 and a model that fits 2*d*-HDMR-GPR using linear combinations of the original variables built with the choices of matrices $\boldsymbol{A}_n$, that we will call 2*d**-HDMR-GPR), each using 500 and 1,000 training points and present the prediction results on the entire dataset of 10,001 points (i.e. all 10,001 points form the test set). Each of these models uses 50 self-consistency cycles to train, and each component function is trained with a GPR using an RBF (radial basis function, which is equivalent to squared exponential) kernel of length scale 0.6 in each dimension and a noise variance of $1 \times 10^{-11}$.[4] The scale down parameters used are $s = 0.1$ and $e = 1$ (Eq. 10). The coordinate selection matrices for the three models are: for the 1*d* model,

$$\boldsymbol{A}_1 = \begin{pmatrix} 1 \\ 0 \\ 0 \end{pmatrix}, \quad \boldsymbol{A}_2 = \begin{pmatrix} 0 \\ 1 \\ 0 \end{pmatrix}, \quad \boldsymbol{A}_3 = \begin{pmatrix} 0 \\ 0 \\ 1 \end{pmatrix} \tag{15}$$

for the 2*d* model,

$$\boldsymbol{A}_{1,2} = \begin{pmatrix} 1 & 0 \\ 0 & 1 \\ 0 & 0 \end{pmatrix}, \quad \boldsymbol{A}_{1,3} = \begin{pmatrix} 1 & 0 \\ 0 & 0 \\ 0 & 1 \end{pmatrix}, \quad \boldsymbol{A}_{2,3} = \begin{pmatrix} 0 & 0 \\ 1 & 0 \\ 0 & 1 \end{pmatrix} \tag{16}$$

and for the 2*d** model,

$$\boldsymbol{A}_{1,23} = \begin{pmatrix} 1 & 0 \\ 0 & 1 \\ 0 & 1 \end{pmatrix}, \quad \boldsymbol{A}_{12,3} = \begin{pmatrix} 1 & 0 \\ 1 & 0 \\ 0 & 1 \end{pmatrix}, \quad \boldsymbol{A}_{13,2} = \begin{pmatrix} 1 & 0 \\ 0 & 1 \\ 1 & 0 \end{pmatrix} \tag{17}$$

The first model implements the approximation $f(\boldsymbol{x}) = \sum_{i=1}^{3} f_i(x_i)$, the second, $f(\boldsymbol{x}) = \sum_{1 \le i < j \le 3} f_{i,j}(x_i, x_j)$, and the last, $f(\boldsymbol{x}) = g_1(x_1, x_2 + x_3) + g_2(x_1 + x_2, x_3) + g_3(x_1 + x_3, x_2)$.

The last model is chosen to illustrate the fact that it may be advantageous to use linear transformations of the original variables.[33]

The RMSE values on the entire dataset with the three models are, respectively, of 459.9, 267.5, and 47.9 with 500 points and 462.5, 259.1, and 52.6 $cm^{-1}$ with 1,000 training point. 500 training points are therefore sufficient to determine first and second order component functions. The reader can verify by changing the corresponding example Python notebook provided in the Supplementary Material that even fewer training points can be used to obtain a similar degree of accuracy. For comparison, the test RMSE of a plain GPR model (which is trivially invoked in the code by using a single matrix $\boldsymbol{A} = \boldsymbol{I}$ or by calling GPR directly from *sklearn* is 1.9 $cm^{-1}$. The RMSE value for the full GPR model is naturally lower because it considers all orders of coupling. The component functions of the trained models and the correlation plots between the target and the fitted values of the potential energy are shown in Figure 1 and Figure 2, respectively. The shapes of the first order component functions correspond to a single potential well nature of the fitted function.[32] The shapes of the second order functions also reflect the single-well nature of the fitted function in the interpolation region of the space. The functions reach large values in the corners of the plot as the corners belong to extrapolation regions where there are no data; they correspond to more than one molecular internal coordinate reaching their maximum or minimum values simultaneously, which results in potential energy values beyond the maximum of 20,000 $cm^{-1}$ included in the data.[32]

As expected, the second-order model performs better than the first order model. One of the things to note is the improvement in RMSE of the third model ($2d^*$) compared to the second. Although they both use $2d$ component functions, the third model performs significantly better. For dimension $d > 1$, the inputs of HDMR component functions can be optimized via taking linear combinations of the original inputs, see Ref. [22, 33-36]. This is one of the reasons why we have implemented RS-HDMR-GPR using matrices to determine the input of component functions, even though in the present implementation the linear combinations are not optimized by the code and need to be defined by the user. In Figure 1, we have also plotted the convergence of the model errors on the training set across the self-consistency cycles. The parameters we choose for Eq. (10) are $s = 0.1, e = 2$, which ensures that $a(c) = 1$ on cycle 25 of 50 cycles, to show the fit behavior after the scaling has been removed. This convergence behavior is typical for all presented models.

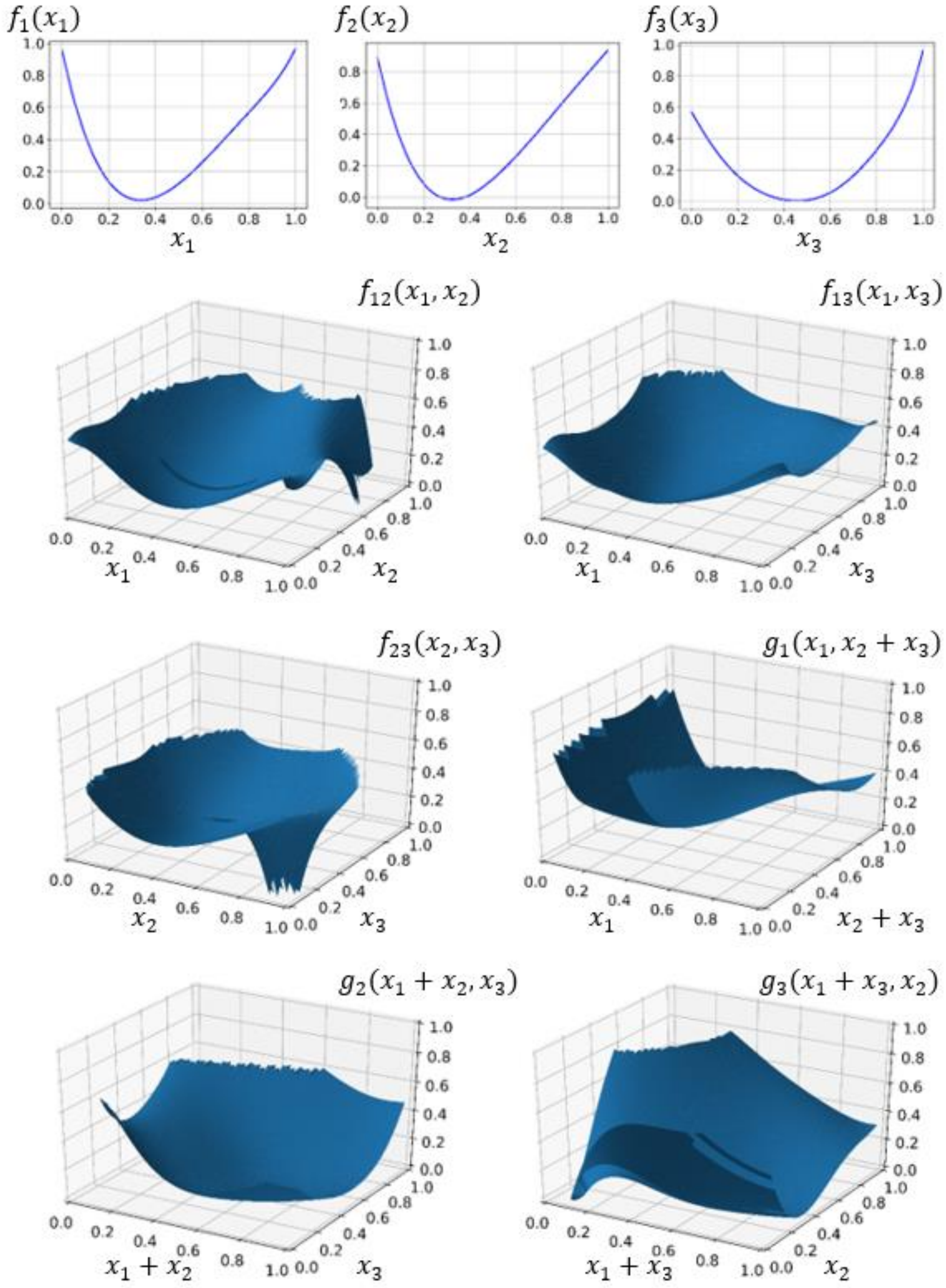

Figure 1. Left to right and top to bottom: the component functions of 1*d*-HDMR-GPR ($f_1(x_1)$, $f_1(x_1)$, $f_1(x_1)$), the component functions of 2*d*-HDMR-GPR ($f_{12}(x_1, x_2)$, $f_{13}(x_1, x_3)$, $f_{23}(x_2, x_3)$), and the component functions of 2*d**-HDMR-GPR ($g_1(x_1, x_2+x_3)$, $g_2(x_1+x_2, x_3)$, $g_3(x_1+x_3, x_2)$) model of the potential energy surface of $H_2O$ fitted to 1,000 training points.

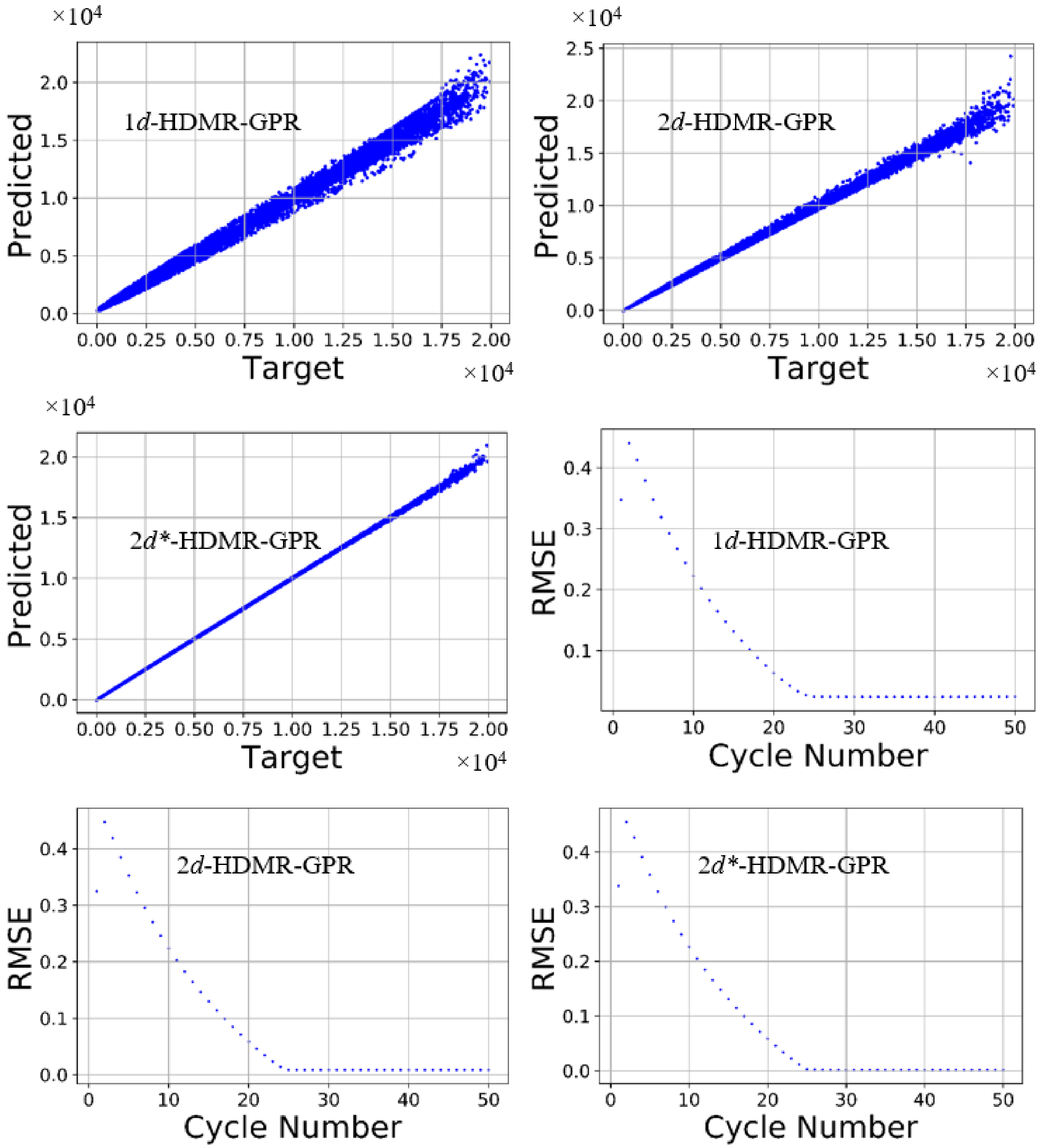

Figure 2. Left to right and top to bottom: predicted vs true values (in $cm^{-1}$) of the $H_2O$ potential energy from 1*d*-HDMR-GPR, 2*d*-HDMR-GPR and 2*d**-HDMR-GPR models fitted to 1,000 training points, and their respective convergence on the training set over cycles in the same order. $a(c)$ of Eq. 10 reaches 1 on cycle $c = 25$. The convergence is given in scaled RMSE (for data scaled to [0,1]). The final RMSE values in $cm^{-1}$ are given in the text.

*3.2 Insight from HDMR-GPR on the example of fitting kinetic energy densities*

In this subsection, we demonstrate how HDMR-GPR can be used to get insight into relative importance of different combinations of variables. We do it by machine learning of kinetic energy densities (KED) from the electron density. This is important, in particular, for the construction of kinetic energy functionals (KEF) for orbital-free density functional theory (OF-DFT).[29] OF-DFT permits near-linear scaled large-scale ab initio modeling and can revolutionize atomistic materials modeling, but the absence of KEFs sufficiently accurate for applied simulations of most materials is hindering its wider use. Machine learning of KEFs has gained significant attention in recent years.[5, 37-41] One way to build a KEF is via a KED model. In it, one typically tries to map the so-called positively definite Kohn-Sham KED, $\tau(\boldsymbol{r}) = \frac{1}{2}\sum_{i=1}|\nabla\psi_i(\boldsymbol{r})|^2$, where $\psi_i(\boldsymbol{r})$ are single-electron orbitals (solutions of the Kohn-Sham equation[42]) and the sum is over all occupied orbitals (here we neglect spin and partial occupancies without loss of generality), as a function of electron density-dependent variables, which may include the density $\rho(\boldsymbol{r}) = \sum_{i=1}|\psi_i(\boldsymbol{r})|^2$ and its powers and derivatives: $\tau(\boldsymbol{r}) = F[\rho(\boldsymbol{r})]$.

One of the key issues is the choice of density-dependent variables in which ML is to be done. It was previously shown that using the terms of the fourth-order gradient expansion,[43] $\tau_{GE4} = \tau_{TF}\left(1 + \frac{5}{27}p + \frac{20}{9}q + \frac{8}{81}q^2 - \frac{1}{9}pq + \frac{8}{243}p^2\right)$, where $p = \frac{|\nabla\rho|^2}{4(3\pi^2)^{2/3}\rho^{8/3}}$ is the scaled squared gradient, $q = \frac{\Delta\rho}{4(3\pi^2)^{2/3}\rho^{5/3}}$ is the scaled Laplacian of the density and $\tau_{TF}(\boldsymbol{r}) = \frac{3}{10}(3\pi^2)^{2/3}\rho^{5/3}(\boldsymbol{r})$ is the Thomas-Fermi KED[44], is advantageous.[5] It was also shown that using the term $\rho(\boldsymbol{r})V_{eff}(\boldsymbol{r})$, where $V_{eff}(\boldsymbol{r})$ is the effective Kohn-Sham potential,[42] can substantially facilitate machine learning.[37] In Ref. [37], Gaussian process regression was performed to fit the KED of magnesium, aluminum, and silicon as a function of seven density-dependent variables $\boldsymbol{x}$ = ($\tau_{TF}$, $\tau_{TF}p$, $\tau_{TF}q$, $\tau_{TF}p^2$, $\tau_{TF}pq$, $\tau_{TF}q^2$, $\rho V_{eff}$). In particular, energy-volume dependences were well reproduced by the GPR model. The dataset was computed with Kohn-Sham DFT (in Abinit[45]) by sampling the conventional standard unit cells of Mg, Al, and Si at equilibrium geometry as well as under compression and extension. In particular, $V_{eff}$, $\rho$, $\nabla\rho$ and $\Delta\rho$ were computed by Abinit and the seven variables ($x_i$ components) were formed from them. Here, we use the same dataset as in Ref. [37] that combines the data for all three materials and all geometries, for a total of about 586,000 points. It was pointed out[5, 37] that the distributions of the values of these variables are extremely uneven. When using a small number of fitting points (we use up to 5,000), this creates a sufficiently low density of sampling so that not all orders of coupling may be reliably recovered.[18, 31] It would be insightful to know which combinations of the seven variables are actually most or at all useful. We show here that HDMR-GPR can do exactly that. The purpose of HDMR-GPR fits performed in this section is not to build a transferrable KED functional; rather, it is to show that HDMR-GPR can be used to obtain insight on the relative importance of different subsets of variables in the fit of a KED. That insight obtained could be used in the building of KEFs with any method (not necessarily machine learning based). It is known that GPR with a squared exponential kernel can be used to determine the importance of different variables whereby the inverse of the length scale parameter determines how relevant an input is; this approach is known as automatic relevance determination (ARD).[4] HDMR-GPR allows determining the importance of different *combinations* of variables. This is a non-trivial extension of capability as the extent of the importance of a variable may be different in different combinations with other variables.

Table 1 shows the results of HDMR-GPR fits of these data with 500, 2000, and 5000 training points randomly drawn from the entire dataset. That is, the largest training set size we used *is less than 1%* of the total number of data, and the test set is the entire dataset of ~586,000 points. Representative correlation plots (of predicted vs target KED values) at each order of HDMR expansion are shown in Supplementary Material (Figure S1). The RMSE between the target KED ($\tau$) and the HDMR-GPR model is reported for the test set. There is some variability from run to run due to the random draw of the training points, but it is relatively small and does not affect the present discussion. Here we use a full HDMR expansion, Eqs. 1, 6. Table 1 shows the results for up to $d$ = 2, as well as those with a plain full-dimensional GPR, and the results for all orders of HMDR are given in the Appendix. The inputs and the target KED were scaled to the range [0, 1]. We used an isotropic squared exponential kernel for each component function, i.e., the length parameter is the same for all variables of a component function, while length parameters are different for different component functions. These calculations were performed with the optimization of the length parameter $l$ of the kernel. The parameter was optimized separately for each component function (CF) in the range (0, $10^5$]. We report in Table 1 the variance of each CF and its optimized length parameter. We observe that for some component functions the variance is substantial, and the length parameter is on the order of 1, while for other CFs, the variance is negligibly small, and the length parameter tends to the upper end of the allowed range. We note that a very large length parameter of a squared exponential kernel of a CF effectively makes the kernel constant over the range of the data. This means that HDMR-GPR did not find that particular CF and that particular combination of variables useful to reduce the error of the representation. We note also that the number of CFs that are found useful increases with the increasing number of the training points. In other words, as the density of sampling increases, more information can be recovered from the data.

As expected, the overall RMSE of the HDMR expansion up to order $d$ is decreasing up to a certain $d$ only, corresponding to the fact that with a given density of sampling only coupling terms up to certain order $d_{max}$ may be recovered.[18, 31] Naturally, the value of $d_{max}$ is becoming higher with the number of training points. For example, with 2,000 training points, there is no advantage using sixth order terms, and only a single fifth order term is found to be useful (see Appendix). Interestingly, no fourth-order CF are found to be useful. Only three third-order terms and less than half of the second-order terms are found to be important. This significantly narrows the choices of combinations of density-dependent variables that should be considered or most likely to be important in a KEF. The information provided by HDMR-GPR is therefore of qualitative nature (i.e. insight) and specifically comes from the combination of HDMR with GPR. Note that already at the second order of HMDR the RMSE is on the same order as in a full-dimensional fit (of which the results are also given in Table 1). With a small number of third and fifth order terms one can obtain smaller errors than in a full-dimensional GPR (see Appendix).

An important consequence of the results shown in Table 1 is that HDMR-GPR can be used to prune the unnecessary component functions from the HDMR expansion. The total number of component functions scales combinatorically with the dimensionality of space and the order $d$. This poses a significant roadblock for practical applications of HDMR in very high-dimensional cases. Pruning the CFs with the help of GPR helps alleviate this problem. In the present example, taking the case of 2,000 training points, there are a total of 119 terms with the full HDMR up to order 5, while the HDMR-GPR based pruning reduces the number of useful terms to only 15.

Table 1. Variances (*var*) of the component functions (CF) and corresponding optimized kernel length parameter values ($l$) of HDMR-GPR of orders $d$ = 1 and 2 when fitting kinetic energy densities with different numbers of training points. The RMSE values on the test set with HDMR-GPR using all terms up to (and including) order $d$ are also given. When $l$ value optimization

resulted in a very large value (100,000 in this case), the corresponding cells are left blank. For comparison, the results with full-dimensional GPR ($d$ = 7) are also given. Data for higher orders $d$ are given in the Appendix. The RMSE values are in atomic units, while variance and length parameter values are with respect to the scaled range of (0, 1].

| Variable combinations | 500 train pts | | 2000 train pts | | 5000 train pts | |
|---|---|---|---|---|---|---|
| | *var*(CF) | $l$ | *var*(CF) | $l$ | *var*(CF) | $l$ |
| RMSE (full-D GPR) | 4.61E-04 | | 3.29E-04 | | 2.53E-04 | |
| $x_1$ ($\tau_{TF}$) | 6.22E-02 | 4.32E-01 | 7.08E-02 | 4.55E-01 | 6.90E-02 | 3.51E-01 |
| $x_2$ ($\tau_{TF}\, p$) | 7.79E-02 | 3.22E-01 | 7.56E-02 | 3.44E-01 | 7.60E-02 | 2.88E-01 |
| $x_3$ ($\tau_{TF}\, q$) | 4.37E-02 | 3.05E+00 | 4.46E-02 | 6.29E-01 | 5.02E-02 | 1.14E-01 |
| $x_4$ ($\tau_{TF}\, p^2$) | 1.62E-10 | | 8.53E-04 | 5.71E+01 | 1.46E-02 | 1.42E+00 |
| $x_5$ ($\tau_{TF}\, pq$) | 1.36E-10 | | 1.23E-10 | | 7.90E-03 | 1.03E+00 |
| $x_6$ ($\tau_{TF}\, q^2$) | 4.68E-11 | | 1.04E-02 | 1.25E-01 | 4.09E-11 | |
| $x_7$ ($\rho\, V_{eff}$) | 1.10E-01 | 4.16E-01 | 1.04E-01 | 3.67E-01 | 1.08E-01 | 2.08E-01 |
| RMSE ($d$ = 1) | 1.06E-03 | | 9.58E-04 | | 9.10E-04 | |
| $x_1$, $x_2$ | 1.54E-10 | | 1.47E-02 | 1.21E+01 | 3.36E-02 | 4.94E-01 |
| $x_1$, $x_3$ | 1.46E-11 | | 1.99E-10 | | 4.51E-04 | 3.33E+02 |
| $x_1$, $x_4$ | 2.33E-10 | | 7.61E-11 | | 3.69E-09 | |
| $x_1$, $x_5$ | 2.57E-10 | | 8.35E-11 | | 3.70E-09 | |
| $x_1$, $x_6$ | 1.19E-10 | | 7.72E-11 | | 3.68E-09 | |
| $x_1$, $x_7$ | 3.64E-11 | | 1.09E-09 | | 3.94E-09 | |
| $x_2$, $x_3$ | 3.29E-02 | 9.31E-01 | 8.10E-03 | 1.36E+01 | 5.09E-03 | 2.16E+01 |
| $x_2$, $x_4$ | 2.15E-10 | | 1.09E-10 | | 3.11E-10 | |
| $x_2$, $x_5$ | 2.60E-10 | | 1.26E-10 | | 2.81E-10 | |
| $x_2$, $x_6$ | 1.17E-10 | | 1.44E-10 | | 1.65E-10 | |
| $x_2$, $x_7$ | 1.53E-10 | | 3.74E-10 | | 6.03E-03 | 3.12E+01 |
| $x_3$, $x_4$ | 2.19E-10 | | 1.94E-10 | | 9.87E-10 | |
| $x_3$, $x_5$ | 2.61E-10 | | 2.09E-10 | | 9.95E-10 | |
| $x_3$, $x_6$ | 1.19E-10 | | 1.52E-10 | | 1.03E-09 | |
| $x_3$, $x_7$ | 3.54E-02 | 8.06E-01 | 2.92E-02 | 1.90E-01 | 3.53E-02 | 9.51E-02 |
| $x_4$, $x_5$ | 4.61E-10 | | 1.01E-02 | 3.30E+00 | 6.70E-10 | |
| $x_4$, $x_6$ | 3.24E-10 | | 2.86E-11 | | 4.76E-10 | |
| $x_4$, $x_7$ | 2.26E-10 | | 1.04E-10 | | 5.08E-10 | |
| $x_5$, $x_6$ | 3.63E-10 | | 3.01E-11 | | 1.38E-10 | |
| $x_7$, $x_7$ | 2.62E-10 | | 1.21E-10 | | 1.78E-10 | |
| $x_6$, $x_7$ | 1.24E-10 | | 9.69E-03 | 3.66E-01 | 1.52E-02 | 2.48E-01 |
| RMSE ($d$ = 2) | 7.75E-04 | | 6.61E-04 | | 4.46E-04 | |

### 3.3 Synthetic datasets

In theory, the $1d$-HDMR imputation of a dataset should be perfect when there is no coupling between the variables. We demonstrate this by creating a synthetic dataset using the function $f(x, y, z) = x + y + z$ for which outputs are known. We generate it using a uniform random generator of 10,000 data entries, for each input variable $x, y, z$ all of whose values belong to the closed interval $[0, 1]$. Our feature matrix $\boldsymbol{X}$ will be a 10,000×3 matrix (where the columns represent the values of $x, y$ and $z$ respectively) and our output vector $\boldsymbol{Y}$ is determined by $f(x, y, z)$. We scale $\boldsymbol{Y}$ so that its components are also within $[0,1]$. Taking 400 rows from the generated dataset, we set 300 aside and randomly introduce missing values. The missing values are introduced so that each row in this set has exactly one missing value, and the missing values are distributed evenly across the columns (i.e. each column has exactly 100 missing values). Using only the remaining 100 points for training a $1d$-HDMR model, we obtain an RMSE of 0.0027 upon prediction using the entire dataset. Because of no coupling, in theory we expect almost perfect imputation on the dataset. Indeed, after imputing the 300 points and retraining on all 400 rows, we obtain an RMSE of 0.0028 on the entire dataset. We point out that these reported RMSE values are subject to minor changes when a different random seed is selected. This is the case throughout the rest of this section. We have fixed the random seed in the provided code, so that the values reported in this document can be generated. All trainings are done with the RBF kernel and the same hyper-parameters (length scale $=$ 0.6, noise variance $=$ $1 \times 10^{-10}$). The correlation plots between the imputed values of each column of the feature space $\boldsymbol{X}$ with the actual values are shown in Figure 3. Their RMSE values are, respectively, 0.00758, 0.00800, and 0.00748.

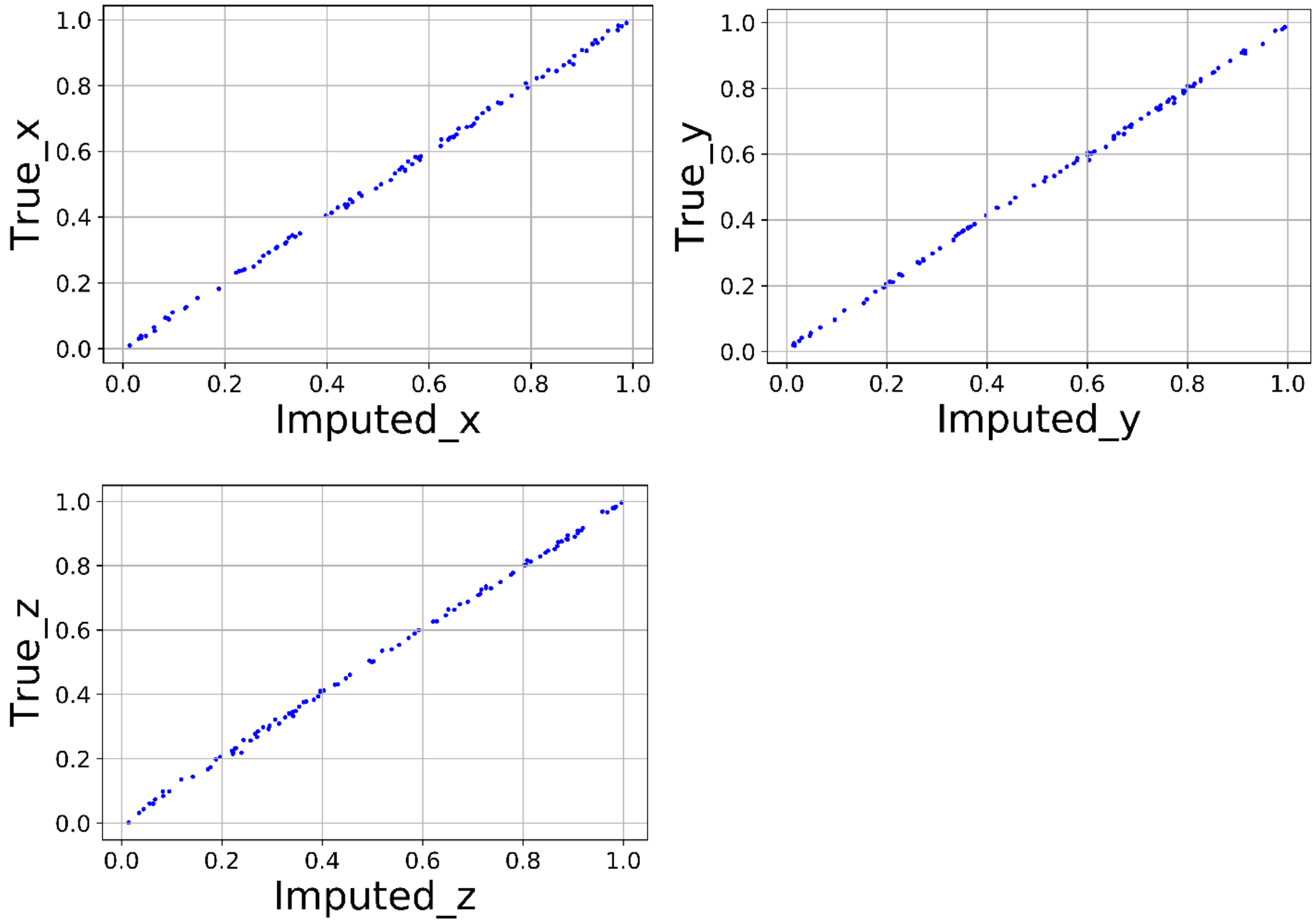

Figure 3. Imputed vs true values of the variables $x, y, z$ of the function $f(x, y, z) = x + y + z$ with 1$d$-HDMR-GPR.

We also point out that for this type of test function, we have also performed tests with data degraded by noise of different levels, increased dimension beyond three, data generated with uneven distributions. All of these factors have an impact on the imputation. To demonstrate this, we have plotted the imputation of the first component ($x_1$) for:

1. A 15-dimensional example, i.e. $f(x_1, x_2, \ldots, x_{15}) = x_1 + x_2 + \cdots + x_{15}$. RMSE: 0.018.
2. The 3$D$ example with noise added to the target having a standard deviation of 0.05. RMSE: 0.049.
3. The 3$D$ examples with regenerated features using an uneven distribution appending 10,000 data points generated from a normal distribution generator with $\mu = 0.1, \sigma = 0.01$ and 5,000 data points from a uniform generator in [0, 1). RMSE: 0.014.

For a fair comparison, all of these are trained with 200 points and imputed on 100 points.

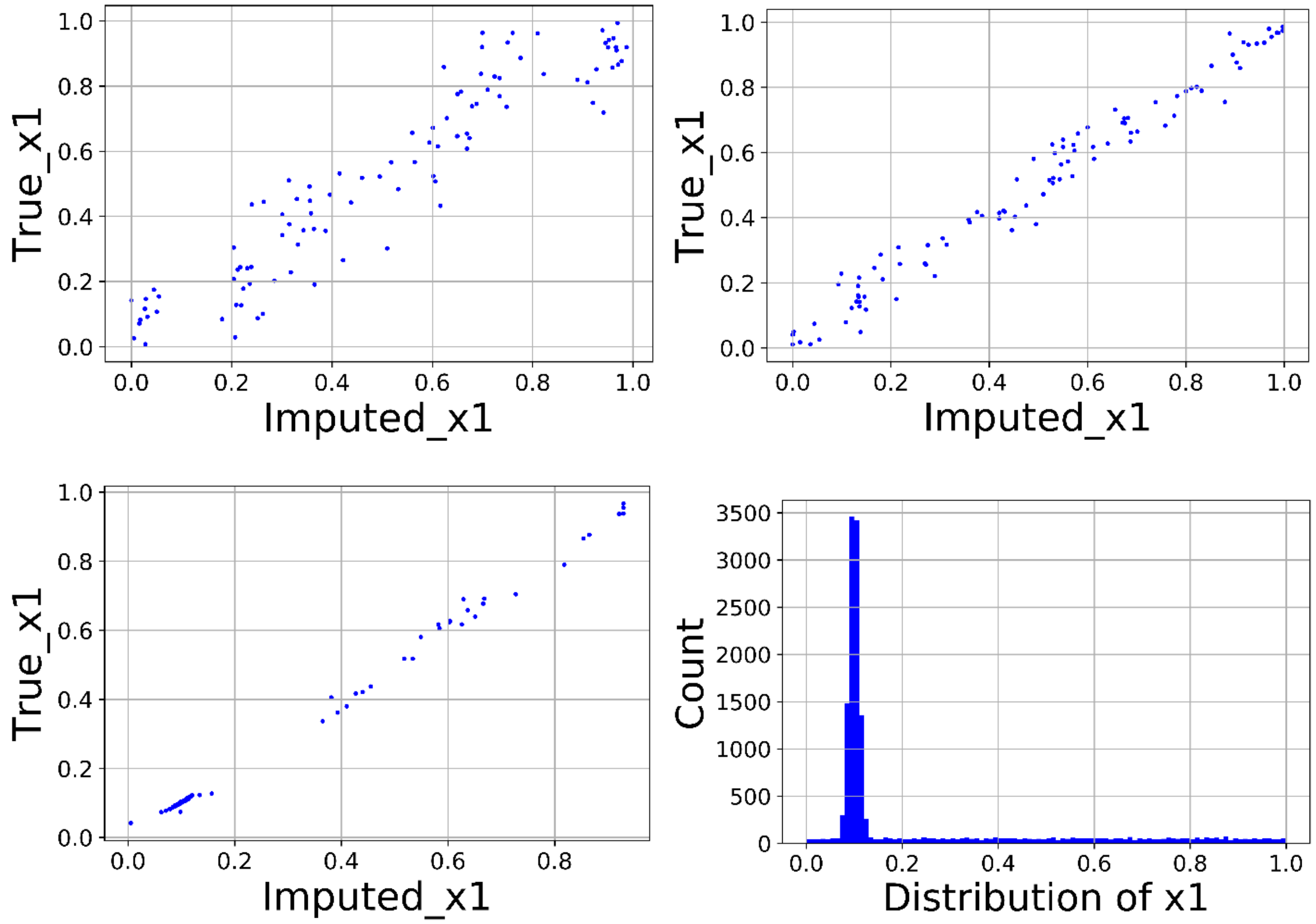

Figure 4. Examples of imputation of missing values of $x_1$ for a 15-dimnesional example (top left), 3-dimensional example with noise (top right), and a 3-dimensional example with uneven distribution (bottom left) as well as the distribution itself (bottom right).

The results of some of the imputations are shown in Figure 4. We see that depending on the extent of the aforementioned distortions to the function, the data distribution, or noise level, the quality of imputation will be affected. The user is therefore advised to control for these factors. All the examples in the remainder of this section, unless otherwise stated, are trained on 100 points and imputed on 100 points for each component function.

To demonstrate the effect of singularities of $f_i^{-1}$ in Eq. 13, we now modify $f(x,y,z)$ to $f(x,y,z) = x^3 + y + z^5$. In this case, we obtain correlation plots shown in Figure 5. Observe that near the origin, the first and third graph imputations are not very accurate. This is due to two reasons. One being that near the origin the functions $y = x^3$ and $y = x^5$ are nearly horizontal. The second reason is we are using 1000 subintervals for imputation, so for a neighborhood around the origin the output values are tightly packed together. Hence, for several values, the inverse within a subinterval can be computed to be the same. We can mitigate this issue by using more subintervals, however, the issue will persist within a smaller neighborhood around the origin where the derivative of $f$ is zero.

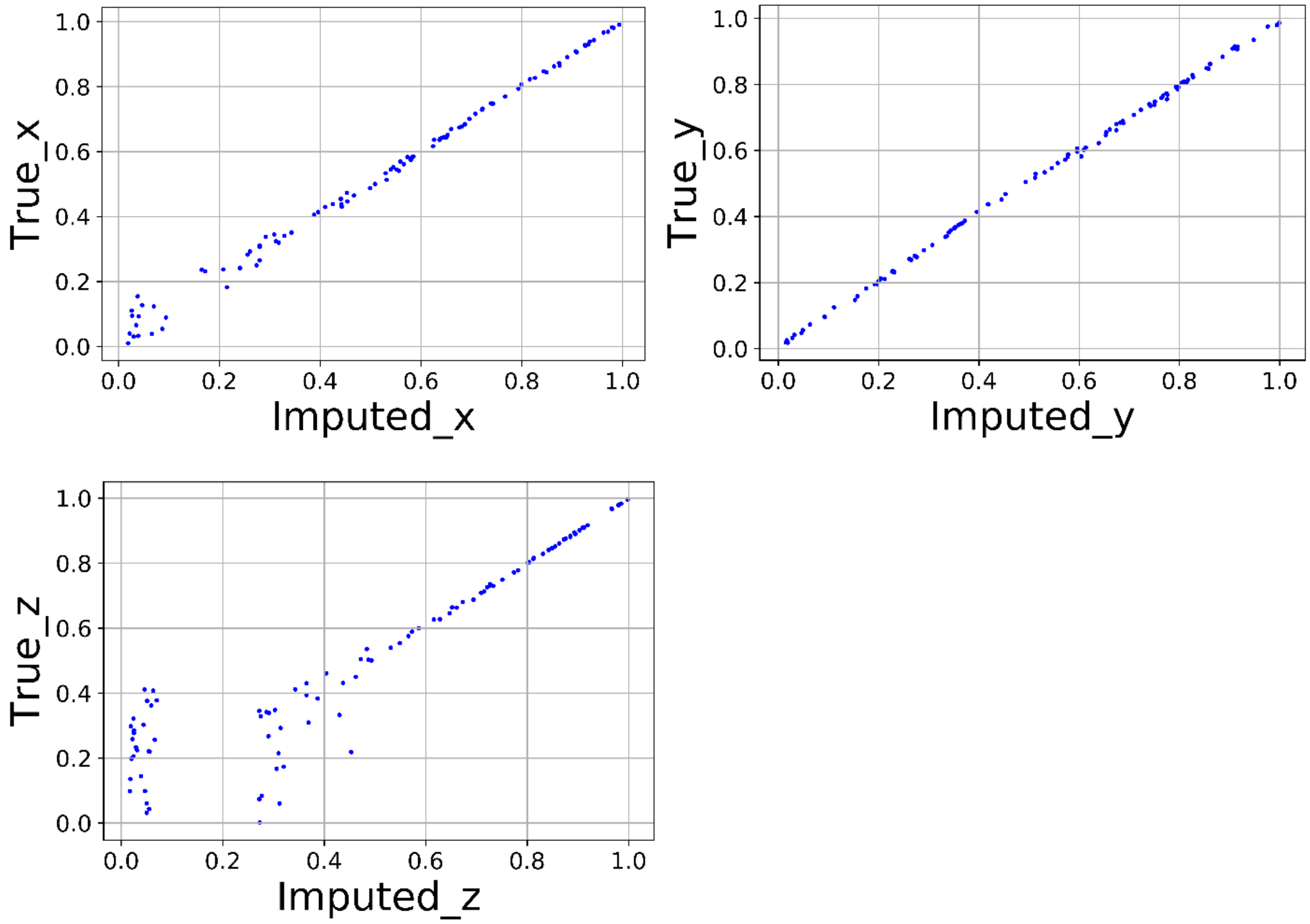

Figure 5. Imputed vs true values of the variables $x$, $y$, $z$ of the function $f(x, y, z) = x^3 + y + z^5$ with 1$d$-HDMR-GPR.

To demonstrate the effect of multivaluedness of $f_i^{-1}$, we use $y = (x - 0.5)^2$ (again on [0, 1]) where the inverse function is not one-to-one. In this case, because the quadratic has two possible values, the correlation plot may look like an "X" or a portion of it. This is expected because the quadratic equation is symmetric about $x = 0.5$. Although there are two possible values in this case, we do point out that among the candidates for the imputation, there is always a correct choice. Consider $f(x, y, z) = 0.5((3.5(x - 0.5))^4 - \left(5.5(x - 0.5)\right)^2 + 1.6) + y + z$. The $x$-component of this function can be quadruple-valued for $x$ in [0, 1]. In Figure 6, we have graphed the possible imputed values with the actual values of the *x*-component, using 100 randomly selected points and setting it to missing.

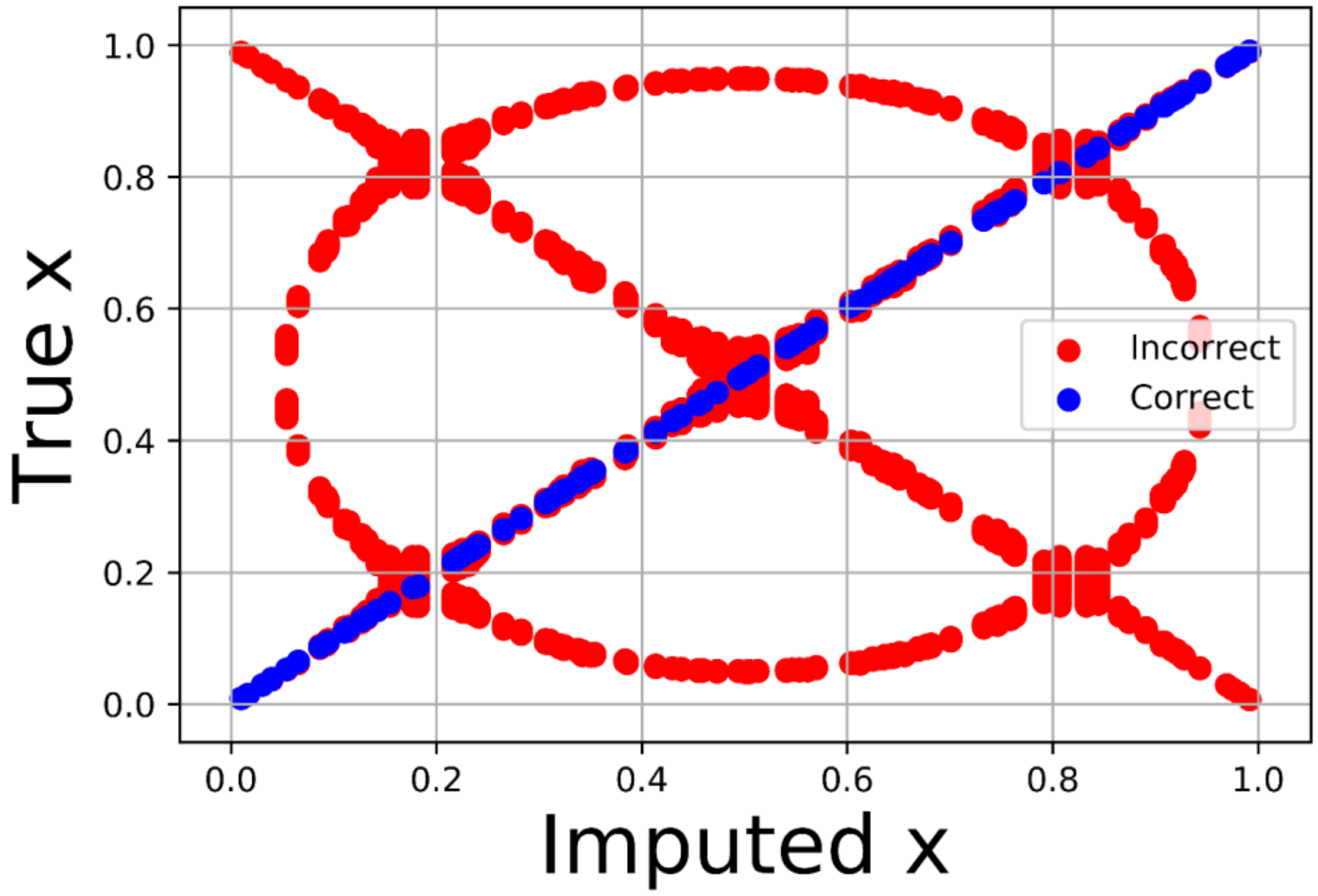

Figure 6. Possible imputed values of $x$ for $f(x,y,z) = 0.5((3.5(x - 0.5))^4 - (5.5(x - 0.5))^2 + 1.6) + y + z$.

The figure shows that the correct choice is always among the candidates provided i.e. HDMR-GPR still significantly narrows down the choices of imputed values even when it cannot unambiguously select the right value. In fact, if the model is accurate, it is guaranteed that the there will always be a correct choice among the candidates given. When a component function has multiple local minimums or is flat at certain regions, the imputation is not accurate because of the error in the 1*d*-HDMR model. It is possible in some cases that the correct value is never considered, due to the fact that the correct candidate maybe be on a flat region of the curve that is at a distance δ (that is the size of the error of the 1*d*-HDMR approximation) away from the value being matched. For example, consider $g(x) = 0.5((3.5(x - 0.5))^4 - (5.5(x - 0.5))^2 + 1.6)$ and Figure 7.

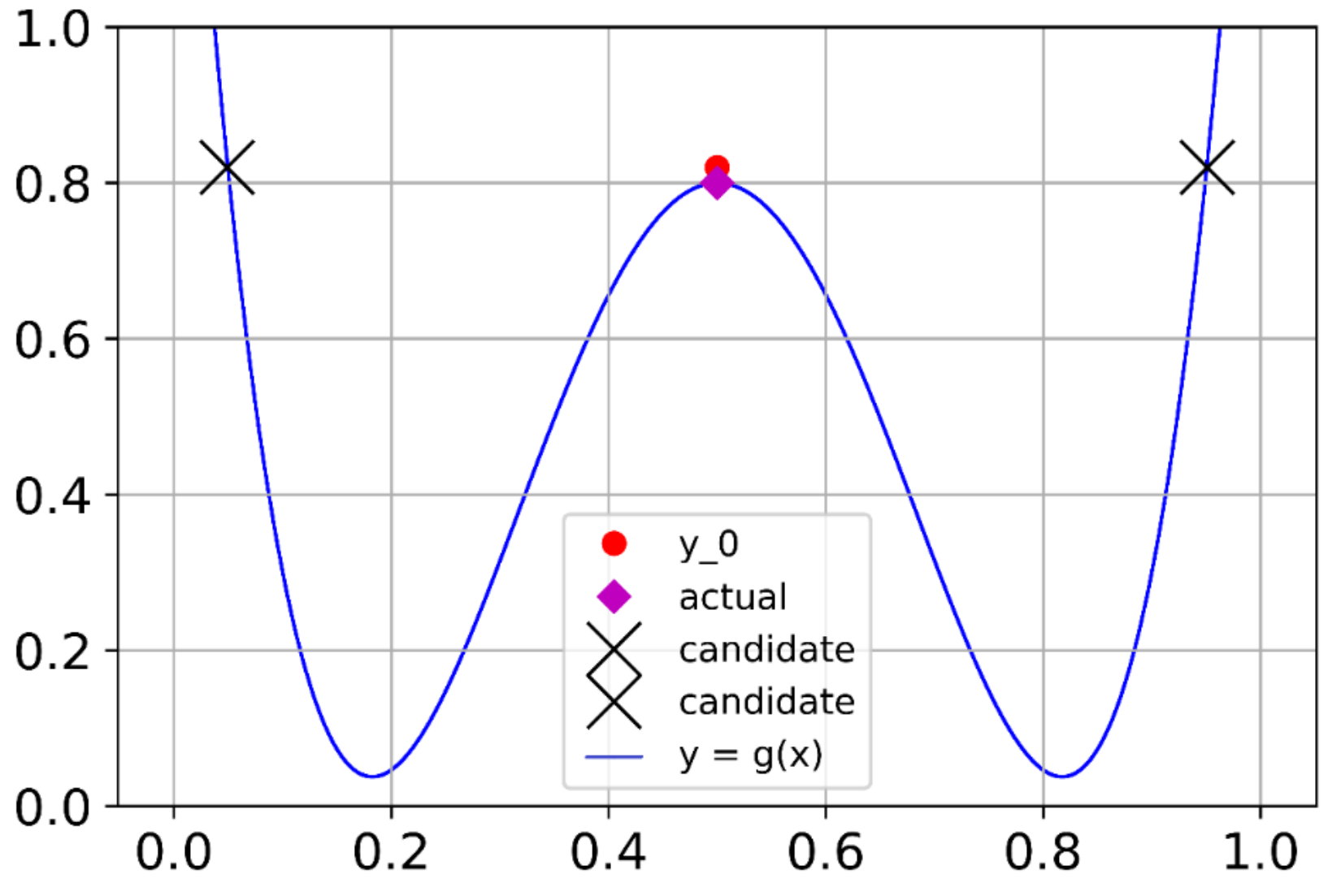

Figure 7. Illustration of a possible effect of the error due to the $1d$-HDMR approximation on imputation. The function being illustrated is $g(x) = 0.5((3.5(x - 0.5))^4 - \left(5.5(x - 0.5)\right)^2 + 1.6)$.

The $y$ coordinate of the red point is the computed output of $g(x)$. By our method of imputation, we could end up with the two points marked 'X's as candidates. However, the true correct imputation would be the purple diamond point. The distance between the red and purple point is due to the error of the $1d$-HDMR model underlying the imputation. From Eq. 13, the computed output of a component function is taken as the difference of the label with the sum of the other component outputs. This implies that the computed output has error at least equal to the error coming from the $1d$-HDMR model. To ensure that we can still obtain the actual value, we can set a threshold distance $\delta$, so that the candidates within that distance will be considered. However, one should be careful as not to set the threshold to be too high to ensure that the number of candidates to consider remains reasonable. The threshold can be set in the code.

The last synthetic example we use is $f(x, y, z) = x + 0.2xy + y + z$ to demonstrate the effect of coupling of the inputs. Here $x$ and $y$ are correlated. The resulting imputations based on $1d$-HDMR, as expected, is less accurate compared to the first example without the $0.2xy$ term. The resulting RMSE values of the imputations are 0.0188, 0.0172, and 0.0217 for $x$, $y$, and $z$ variable, respectively. The correlation plots are shown in Figure 8. Imputation is reasonable even for coupled variables as long as coupling terms are relatively small (small enough for a reasonable 1$d$-HDMR approximation). We point out that the variable $z$'s imputation accuracy is also affected because of the higher error from the model. The imputation error for any variable will always be at least as large as the error from the model.

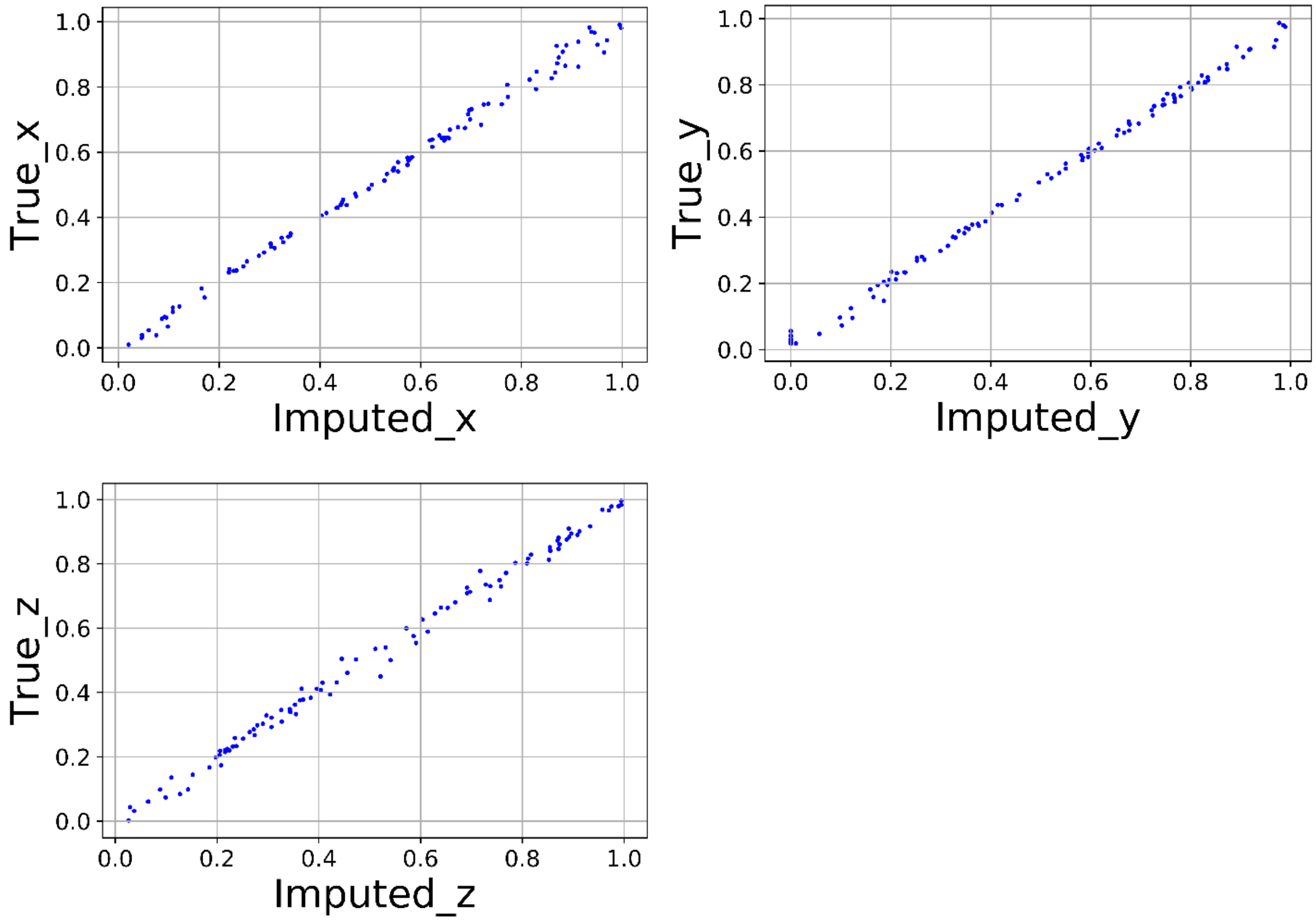

Figure 8. Imputed vs true values of the variables $x$, $y$, $z$ of the function $f(x, y, z) = x + 0.2xy + y + z$ with 1$d$-HDMR-GPR.

### 3.4 *Example from quantitative finance*

The algorithm RS-HDMR-GPR is general and can be very useful in many applied domains of science, technology, and business. Stock market indexes prediction represents a stringent test for making predictions and imputation. Indeed, stock price time series are often characterized by a chaotic and nonlinear behavior which makes the prediction a challenging task. We demonstrate in this section the use of the HDMR-GPR method for learning the dependence of S&P 500 index on other financial variables and on imputing missing values in this type of real datasets.

The data were collected for the period of 2001-01-01 to 2020-07-31. To see how the value of S&P 500 is related to other major world indexes and selected currencies and commodities, we include them as features. The period we consider includes bear and bull markets including some major crashes as well as periods with relatively high and low interest rates. As different data series had different missing dates (weekends, holidays), the total dataset had incomplete rows in which the missing values were interpolated with splines. We then scaled the data to [0, 1] to simplify the hyper-parameter tuning of the model. The final data consist of 3927 rows and 15 columns (the last column being the value of S&P 500 index). The first 14 columns represent the feature which includes the time series major financial indexes DAX, Nikkei 225, Nasdaq composite, aggregate US-traded bond index (ticker symbol AGG, ticker symbols refer to the US markets if not states otherwise), and volatility index (VIX); macroeconomic variables money supply M2 and short term rates (exemplified by the 3-months US Treasury bill rate, ticker symbol IRX), and commodities prices such as gold (proxied by the gold bullion ETF, ticker symbol GLD) and oil (proxied by a crude oil ETF traded on TSX, ticker symbol HUC) as well as major currencies the Japanese yen (JPY), the Euro (EUR), the Chinese yuan (CNY), and the Canadian dollar (CAD). The IQ Real Return ETF (ticker symbol CPI) is used to include inflation information. The vector of features is $\boldsymbol{X}$ = (CPI, ^IRX, GLD, HUC.TO, AGG, ^VIX, ^N225, ^GDAXI, ^IXIC, CAD, JPY, EUR, CNY, M2) (ticker symbols prefixed with ^ are indices) and the target $Y$ is the S&P500 index. We denote by $\boldsymbol{x} = (x_1, x_2, \ldots, x_{14})$ the feature vector with the elements scaled to [0, 1].

After optimizing the model hyper-parameters (the width of the kernel and the noise), the length scale of the isotropic squared exponential covariance function was chosen to be 0.6 and the noise variance is in the range of $10^{-6}$. Each model uses 50 self-consistency cycles to train. Refitting the financial dataset with RS-HDMR-GPR represents a good test as the dimensionality is relatively high and the data come from the real world and include a random component. The RMSE values on the entire dataset of first, second, and full dimensional RS-HDMR-GPR models are, respectively, 31.6, 14.7, and 16.7. The correlation plots of the predicted vs actual values of the S&P 500 index are shown in Figure 9 for the 1$d$-HMDR, 2$d$-HMDR, and full-D GPR models. One can see that the first order HDMR already obtains a good accuracy.

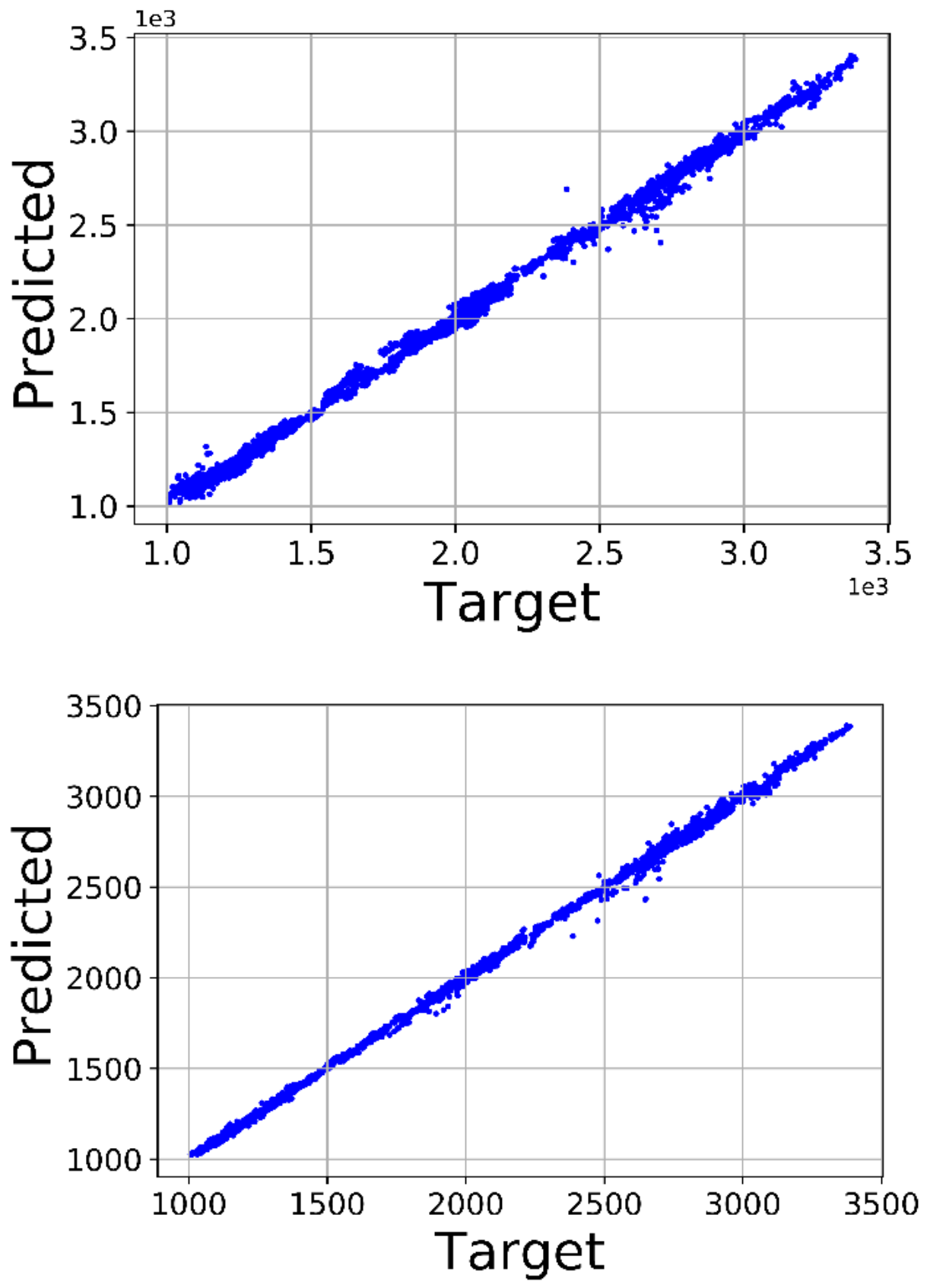

Figure 9. Predicted vs actual values of the S&P500 index with the 1$d$-HDMR-GPR, 2$d$-HDMR-GPR model and the full-$d$-GPR model.

We now look at the imputation performance for this dataset. To test how well the imputation performs on each column, we pick the column that we want to test and assign randomly 100 values along that column as missing. As in section 3.3, we need to ensure that only one entry per row is missing. Figure 10 shows examples of two component functions chosen to illustrate cases when the imputation is reliable and unreliable (due to nearly flat regions in the component function). We have chosen the best and the worst (for imputation) variable / component function to illustrate the performance of the imputation. All component functions are shown in the Supplementary Material (Figure S2, with imputations shown in Figure S3). The blue points in the imputation plots in Figure 10 illustrate the quality of imputation if we selected the best candidates from the choices provided by Eq. 14. We remind the user that our method does not provide a means of selecting the correct choice but computes a list of possible values and in general guarantees most of the time the correct (closest) choice will be within the list. The blue dots in the picture represents the closest choice to the actual value, while the red dots show other potential candidates. The threshold argument we set for distance δ is 0 (i.e. we take into account the case discussed in section 3.3).

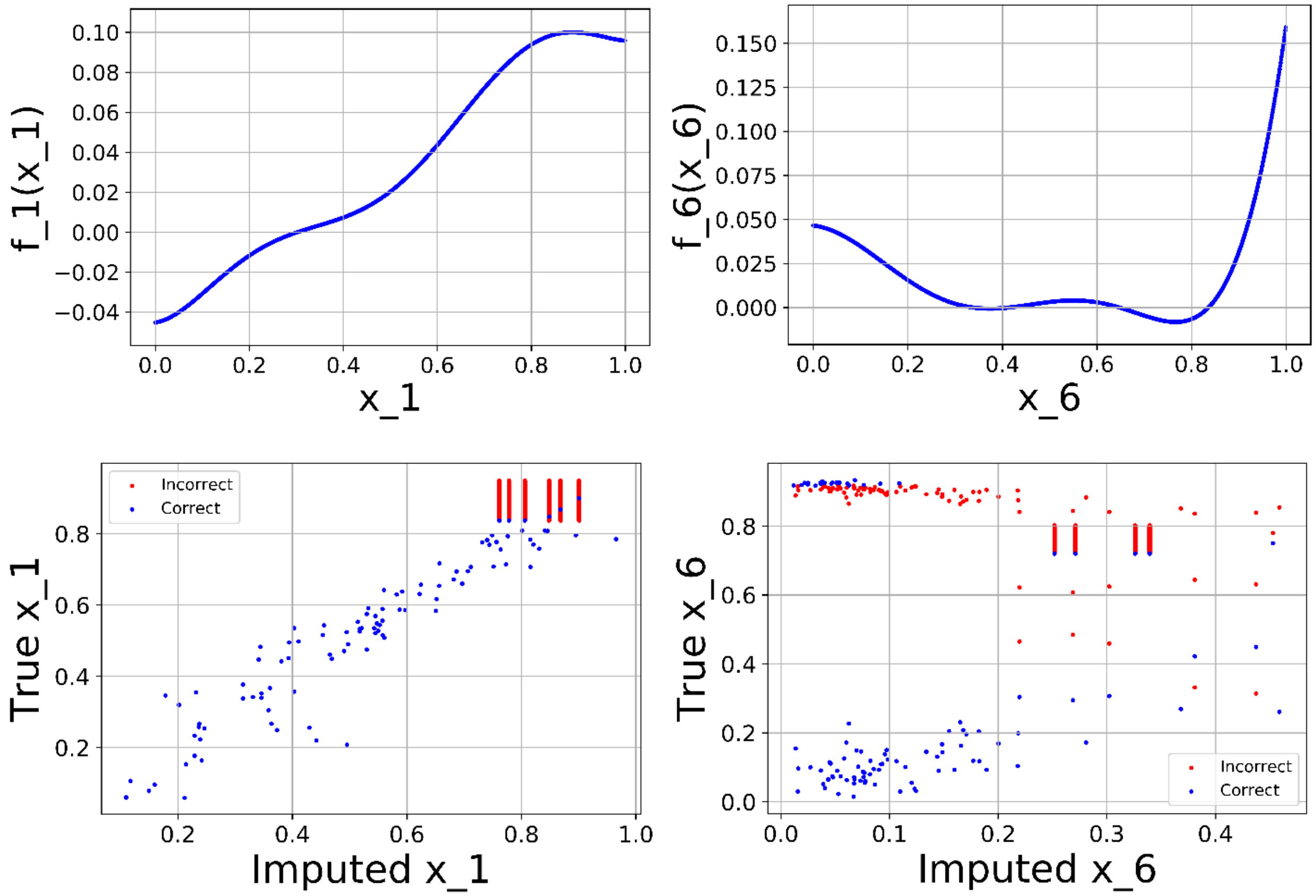

Figure 10. Top row: examples of component functions. Bottom row: imputed vs true values of the corresponding variables.

Assuming the best possible imputation, the RMSE for the worst ($x_6$ corresponding to ^VIX) and the best ($x_1$ corresponding to CPI) cases are, respectively, 0.072 and 0.407. Based on the graphs of the component functions, we expected this to be the case: the more monotonic and less flat curve results in better imputation compared to a flatter and multi-valued function (see section 3.3). The higher errors for the imputation are also due to errors from 1*d*-HDMR model, and the distributions of the features are also very uneven which are also factors we explored with the synthetic dataset in section 3.3. Nevertheless, it is in general more precise than conventional methods of imputing such as using the mean of the distribution.

# 4 Conclusions

We developed the Random Sampling High Dimensional Model Representation Gaussian Process Regression (RS-HDMR-GPR) method that builds approximations to multivariate functions as a sum of lower-dimensional terms of chosen dimensionality and its implementation in a Python module. The approach is based on the HDMR idea which is expanded to non-orthogonal component functions. The formal expansion over orders of coupling (i.e. including all terms from the 1$^{st}$ until the $d$-th order) can be used as well as a version where only functions of a chosen dimensionality $d$ are used. Our implementation allows combining at will component functions of different dimensionality, including using linear combinations of the original variables, as well as imputation of missing data values.

The component functions are built with Gaussian process regressions. The use of GPR avoids the calculation of multidimensional integrals which are needed in the standard RS-HDMR approach. Using lower-dimensional functions allows easier construction of these functions, in particular, from fewer data. This is particularly important for a method like GPR which scales rather unfavorably with the number of data points. This representation also simplifies the use of the function in applications, in particular, its integration.

An important advantage of HDMR-GPR, which we demonstrated when fitting kinetic energy densities, is its ability to identify important and unimportant combinations of variables and component functions when fitting a given dataset. This is in particular important for gaining insight into the nature of the function being fitted but also in view of the combinatorial scaling of the number of terms of HDMR when all terms are used. RS-HDMR-GPR effectively prunes the set of component functions, thus alleviating this important issue.

The HDMR structure allows for imputation of missing values of variables. Specifically, when only one variable is missing out of $D$ (in however many data points), and when 1$d$-HDMR approximation is reasonable, the imputation can be reliable. In most physical systems, the importance of coupling terms tends to decay rapidly with the order if coupling, so this is expected to be relevant in many practical applications. The imputation quality depends not only on the quality of the 1$d$-HDMR model but also on the shape of the component functions: in the absence of flat regions the imputation can be very reliable, while the presence of such regions, compounded by the finite accuracy of the first order model, can make imputation difficult. Nevertheless, the 1$d$-HDMR model significantly narrows down possible choices of missing values and should help address this important problem for which substantial generic but accurate approaches are still lacking.

A code implementing the method is provided in Supplementary Material and is also available at https://github.com/owen-ren0003/rshdmrgpr . Our software relies on the GPR functionality of *sklearn*. The HDMR-GPR approach can of course also be implemented with other GPR engines, for example those which specifically aim at working with large data sets, such as sparse GPR approaches.[46]

## 5 Acknowledgements

This work was partially supported in part by National Sciences and Engineering Research Council (NSERC) of Canada and Compute Canada. O. R. is supported by the Mitacs Elevate program.

## 6 Appendix

Variances (*var*) of the component functions (CF) and corresponding optimized kernel length parameter values (*l*) of HDMR-GPR of all orders when fitting kinetic energy densities with different numbers of training points. The EMSE values on the test set with HDMR-GPR using all terms up to (and including) order *d* are also given. For comparison, the results with full-dimensional GPR are also given. When *l* value optimization resulted in a very large value (100,000 in this case), the corresponding cells are left blank. Data for orders *d* = 1 and 2 are also given in Table 1 above. The RMSE values are in atomic units, while variance and length parameter values are with respect to the scaled range of [0, 1].

| Variable combinations | 500 train pts | | 2000 train pts | | 5000 train pts | |
|---|---|---|---|---|---|---|
| | *var*(CF) | *l* | *var*(CF) | *l* | *var*(CF) | *l* |
| RMSE (full-D GPR) | 4.61E-04 | | 3.29E-04 | | 2.53E-04 | |
| x1 | 6.22E-02 | 4.32E-01 | 7.08E-02 | 4.55E-01 | 6.90E-02 | 3.51E-01 |
| x2 | 7.79E-02 | 3.22E-01 | 7.56E-02 | 3.44E-01 | 7.60E-02 | 2.88E-01 |
| x3 | 4.37E-02 | 3.05E+00 | 4.46E-02 | 6.29E-01 | 5.02E-02 | 1.14E-01 |
| x4 | 1.62E-10 | | 8.53E-04 | 5.71E+01 | 1.46E-02 | 1.42E+00 |
| x5 | 1.36E-10 | | 1.23E-10 | | 7.90E-03 | 1.03E+00 |
| x6 | 4.68E-11 | 4.16E-01 | 1.04E-02 | 1.25E-01 | 4.09E-11 | |
| x7 | 1.10E-01 | | 1.04E-01 | 3.67E-01 | 1.08E-01 | 2.08E-01 |
| RMSE ($d$ = 1) | 1.06E-03 | | 9.58E-04 | | 9.10E-04 | |
| x1,x2 | 1.54E-10 | | 1.47E-02 | 1.21E+01 | 3.36E-02 | 4.94E-01 |
| x1,x3 | 1.46E-11 | | 1.99E-10 | | 4.51E-04 | 3.33E+02 |
| x1,x4 | 2.33E-10 | | 7.61E-11 | | 3.69E-09 | |
| x1,x5 | 2.57E-10 | | 8.35E-11 | | 3.70E-09 | |
| x1,x6 | 1.19E-10 | | 7.72E-11 | | 3.68E-09 | |
| x1,x7 | 3.64E-11 | | 1.09E-09 | | 3.94E-09 | |
| x2,x3 | 3.29E-02 | 9.31E-01 | 8.10E-03 | 1.36E+01 | 5.09E-03 | 2.16E+01 |
| x2,x4 | 2.15E-10 | | 1.09E-10 | | 3.11E-10 | |
| x2,x5 | 2.60E-10 | | 1.26E-10 | | 2.81E-10 | |
| x2,x6 | 1.17E-10 | | 1.44E-10 | | 1.65E-10 | |
| x2,x7 | 1.53E-10 | | 3.74E-10 | | 6.03E-03 | 3.12E+01 |
| x3,x4 | 2.19E-10 | | 1.94E-10 | | 9.87E-10 | |
| x3,x5 | 2.61E-10 | | 2.09E-10 | | 9.95E-10 | |
| x3,x6 | 1.19E-10 | | 1.52E-10 | | 1.03E-09 | |
| x3,x7 | 3.54E-02 | 8.06E-01 | 2.92E-02 | 1.90E-01 | 3.53E-02 | 9.51E-02 |
| x4,x5 | 4.61E-10 | | 1.01E-02 | 3.30E+00 | 6.70E-10 | |
| x4,x6 | 3.24E-10 | | 2.86E-11 | | 4.76E-10 | |
| x4,x7 | 2.26E-10 | | 1.04E-10 | | 5.08E-10 | |
| x5,x6 | 3.63E-10 | | 3.01E-11 | | 1.38E-10 | |
| x5,x7 | 2.62E-10 | | 1.21E-10 | | 1.78E-10 | |
| x6,x7 | 1.24E-10 | | 9.69E-03 | 3.66E-01 | 1.52E-02 | 2.48E-01 |
| RMSE ($d$ = 2) | 7.75E-04 | | 6.61E-04 | | 4.46E-04 | |
| x1,x2,x3 | 1.80E-11 | | 1.27E-02 | 9.31E-01 | 5.25E-11 | |
| x1,x2,x4 | 1.45E-10 | | 5.34E-10 | | 3.45E-10 | |
| x1,x2,x5 | 1.35E-10 | | 5.03E-10 | | 1.65E-10 | |
| x1,x2,x6 | 6.04E-11 | | 1.77E-10 | | 1.57E-10 | |
| x1,x2,x7 | 2.92E-02 | 4.19E-01 | 1.85E-02 | 3.24E-01 | 1.91E-10 | |
| x1,x3,x4 | 9.26E-11 | | 6.89E-10 | | 4.20E-10 | |
| x1,x3,x5 | 9.30E-11 | | 7.06E-10 | | 9.00E-11 | |
| x1,x3,x6 | 9.94E-11 | | 3.52E-10 | | 9.74E-11 | |
| x1,x3,x7 | 1.21E-10 | | 1.92E-10 | | 1.51E-02 | 2.27E-01 |
| x1,x4,x5 | 1.86E-10 | | 1.13E-09 | | 6.64E-10 | |
| x1,x4,x6 | 1.22E-10 | | 7.61E-10 | | 5.56E-10 | |
| x1,x4,x7 | 9.54E-11 | | 5.66E-10 | | 5.26E-10 | |
| x1,x5,x6 | 1.14E-10 | | 8.05E-10 | | 1.89E-10 | |
| x1,x5,x7 | 8.54E-11 | | 5.88E-10 | | 1.52E-10 | |
| x1,x6,x7 | 3.08E-11 | | 2.37E-10 | | 4.93E-11 | |
| x2,x3,x4 | 9.27E-11 | | 6.79E-10 | | 6.52E-03 | 1.11E+01 |
| x2,x3,x5 | 8.99E-11 | | 6.96E-10 | | 4.47E-10 | |
| x2,x3,x6 | 9.03E-11 | | 3.42E-10 | | 3.01E-10 | |
| x2,x3,x7 | 2.48E-02 | 1.87E+00 | 1.36E-02 | 3.22E+00 | 1.09E-02 | 4.00E+00 |
| x2,x4,x5 | 2.40E-10 | | 1.02E-09 | | 2.04E-10 | |
| x2,x4,x6 | 1.63E-10 | | 6.81E-10 | | 2.49E-10 | |
| x2,x4,x7 | 1.54E-10 | | 4.72E-10 | | 2.91E-10 | |
| x2,x5,x6 | 1.55E-10 | | 7.32E-10 | | 2.26E-10 | |
| x2,x5,x7 | 1.45E-10 | | 5.00E-10 | | 1.82E-10 | |
| x2,x6,x7 | 7.12E-11 | | 2.03E-10 | | 1.54E-10 | |
| x3,x4,x5 | 1.99E-10 | | 1.10E-09 | | 2.94E-10 | |
| x3,x4,x6 | 1.24E-10 | | 7.54E-10 | | 3.37E-10 | |
| x3,x4,x7 | 1.03E-10 | | 5.45E-10 | | 3.68E-10 | |
| x3,x5,x6 | 1.21E-10 | | 7.97E-10 | | 1.73E-10 | |
| x3,x5,x7 | 9.88E-11 | | 5.64E-10 | | 9.99E-11 | |
| x3,x6,x7 | 2.98E-11 | | 2.40E-10 | | 8.83E-11 | |
| x4,x5,x6 | 2.56E-10 | | 1.25E-09 | | 1.99E-10 | |
| x4,x5,x7 | 2.44E-10 | | 1.03E-09 | | 2.31E-10 | |
| x4,x6,x7 | 1.67E-10 | | 6.89E-10 | | 2.71E-10 | |
| x5,x6,x7 | 1.62E-10 | | 7.33E-10 | | 2.37E-10 | |
| RMSE (*d*=3) | 3.60E-04 | | 3.56E-04 | | 2.72E-04 | |
| x1,x2,x3,x4 | 4.09E-10 | | 2.01E-09 | | 1.32E-09 | |
| x1,x2,x3,x5 | 3.77E-10 | | 2.11E-09 | | 4.58E-10 | |
| x1,x2,x3,x6 | 1.07E-10 | | 8.92E-10 | | 3.84E-10 | |
| x1,x2,x3,x7 | 1.52E-10 | | 1.57E-10 | | 4.89E-10 | |
| x1,x2,x4,x5 | 9.28E-10 | | 3.82E-09 | | 8.39E-10 | |
| x1,x2,x4,x6 | 6.40E-10 | | 2.54E-09 | | 9.76E-10 | |
| x1,x2,x4,x7 | 6.42E-10 | | 1.78E-09 | | 1.23E-09 | |
| x1,x2,x5,x6 | 6.11E-10 | | 2.73E-09 | | 9.74E-10 | |
| x1,x2,x5,x7 | 6.12E-10 | | 1.87E-09 | | 9.68E-10 | |
| x1,x2,x6,x7 | 3.67E-10 | | 7.97E-10 | | 8.77E-10 | |
| x1,x3,x4,x5 | 7.55E-10 | | 4.10E-09 | | 1.09E-09 | |
| x1,x3,x4,x6 | 4.65E-10 | | 2.81E-09 | | 1.24E-09 | |
| x1,x3,x4,x7 | 4.38E-10 | | 2.04E-09 | | 1.44E-09 | |
| x1,x3,x5,x6 | 4.57E-10 | | 2.97E-09 | | 7.04E-10 | |
| x1,x3,x5,x7 | 4.27E-10 | | 2.10E-09 | | 6.78E-10 | |
| x1,x3,x6,x7 | 2.02E-10 | | 9.08E-10 | | 6.26E-10 | |
| x1,x4,x5,x6 | 9.87E-10 | | 4.66E-09 | | 8.52E-10 | |
| x1,x4,x5,x7 | 9.70E-10 | | 3.84E-09 | | 1.08E-09 | |
| x1,x4,x6,x7 | 6.90E-10 | | 2.59E-09 | | 1.19E-09 | |
| x1,x5,x6,x7 | 6.76E-10 | | 2.75E-09 | | 1.15E-09 | |
| x2,x3,x4,x5 | 8.01E-10 | | 4.01E-09 | | 9.76E-10 | |
| x2,x3,x4,x6 | 5.24E-10 | | 2.73E-09 | | 1.14E-09 | |
| x2,x3,x4,x7 | 4.45E-10 | | 1.93E-09 | | 1.25E-09 | |
| x2,x3,x5,x6 | 4.99E-10 | | 2.91E-09 | | 7.26E-10 | |
| x2,x3,x5,x7 | 4.12E-10 | | 2.03E-09 | | 4.50E-10 | |
| x2,x3,x6,x7 | 1.26E-10 | | 8.28E-10 | | 3.67E-10 | |
| x2,x4,x5,x6 | 1.01E-09 | | 4.58E-09 | | 6.15E-10 | |
| x2,x4,x5,x7 | 9.70E-10 | | 3.75E-09 | | 7.27E-10 | |
| x2,x4,x6,x7 | 6.82E-10 | | 2.48E-09 | | 8.79E-10 | |
| x2,x5,x6,x7 | 6.53E-10 | | 2.67E-09 | | 9.69E-10 | |
| x3,x4,x5,x6 | 8.62E-10 | | 4.85E-09 | | 9.39E-10 | |
| x3,x4,x5,x7 | 7.92E-10 | | 4.02E-09 | | 1.00E-09 | |
| x3,x4,x6,x7 | 5.00E-10 | | 2.74E-09 | | 1.16E-09 | |
| x3,x5,x6,x7 | 4.90E-10 | | 2.90E-09 | | 6.98E-10 | |
| x4,x5,x6,x7 | 1.03E-09 | | 4.59E-09 | | 7.41E-10 | |
| RMSE (*d*=4) | 3.60E-04 | | 3.56E-04 | | 2.72E-04 | |
| x1,x2,x3,x4,x5 | 1.87E-11 | | 6.24E-10 | | 1.13E-09 | |
| x1,x2,x3,x4,x6 | 2.06E-11 | | 6.30E-10 | | 1.12E-09 | |
| x1,x2,x3,x4,x7 | 6.07E-03 | 3.49E-01 | 6.24E-10 | | 1.13E-09 | |
| x1,x2,x3,x5,x6 | 2.41E-11 | | 6.29E-10 | | 1.12E-09 | |
| x1,x2,x3,x5,x7 | 4.22E-11 | | 6.24E-10 | | 1.13E-09 | |
| x1,x2,x3,x6,x7 | 4.31E-11 | | 6.29E-10 | | 1.12E-09 | |
| x1,x2,x4,x5,x6 | 2.61E-11 | | 6.27E-10 | | 1.12E-09 | |
| x1,x2,x4,x5,x7 | 4.13E-11 | | 6.22E-10 | | 1.13E-09 | |
| x1,x2,x4,x6,x7 | 4.23E-11 | | 6.27E-10 | | 1.12E-09 | |
| x1,x2,x5,x6,x7 | 4.29E-11 | | 6.26E-10 | | 1.12E-09 | |
| x1,x3,x4,x5,x6 | 1.97E-11 | | 2.16E-11 | | 2.20E-11 | |
| x1,x3,x4,x5,x7 | 3.99E-11 | | 5.94E-03 | 1.94E-01 | 5.23E-03 | 1.31E-01 |
| x1,x3,x4,x6,x7 | 4.09E-11 | | 2.38E-11 | | 3.15E-11 | |
| x1,x3,x5,x6,x7 | 4.09E-11 | | 2.31E-11 | | 3.27E-11 | |
| x1,x4,x5,x6,x7 | 4.00E-11 | | 1.93E-11 | | 3.60E-11 | |
| x2,x3,x4,x5,x6 | 2.36E-11 | | 6.32E-10 | | 1.12E-09 | |
| x2,x3,x4,x5,x7 | 3.31E-11 | | 6.26E-10 | | 1.13E-09 | |
| x2,x3,x4,x6,x7 | 3.43E-11 | | 6.32E-10 | | 1.12E-09 | |
| x2,x3,x5,x6,x7 | 3.53E-11 | | 6.31E-10 | | 1.12E-09 | |
| x2,x4,x5,x6,x7 | 3.51E-11 | | 6.29E-10 | | 1.12E-09 | |
| x3,x4,x5,x6,x7 | 3.23E-11 | | 1.90E-11 | | 3.35E-11 | |
| RMSE (*d*=5) | 4.71E-04 | | 2.86E-04 | | 1.73E-04 | |
| x1,x2,x3,x4,x5,x6 | 9.67E-11 | | 2.36E-09 | | 4.17E-09 | |
| x1,x2,x3,x4,x5,x7 | 1.58E-10 | | 2.33E-09 | | 4.20E-09 | |
| x1,x2,x3,x4,x6,x7 | 1.62E-10 | | 2.36E-09 | | 4.19E-09 | |
| x1,x2,x3,x5,x6,x7 | 1.64E-10 | | 2.35E-09 | | 4.19E-09 | |
| x1,x2,x4,x5,x6,x7 | 1.63E-10 | | 2.35E-09 | | 4.18E-09 | |
| x1,x3,x4,x5,x6,x7 | 1.54E-10 | | 8.75E-11 | | 1.33E-10 | |
| x2,x3,x4,x5,x6,x7 | 1.34E-10 | | 2.36E-09 | | 4.18E-09 | |
| RMSE (*d*=6) | 4.71E-04 | | 2.86E-04 | | 1.73E-04 | |
| x1,x2,x3,x4,x5,x6,x7 | 3.48E-03 | 1.60E-01 | 2.54E-03 | 1.43E-01 | 1.11E-08 | |
| RMSE (*d*=7) | 9.07E-04 | | 4.65E-04 | | 1.73E-04 | |

# Supplementary Material

# Random Sampling High Dimensional Model Representation Gaussian Process Regression (RS-HDMR-GPR) for representing multidimensional functions with machine-learned lower-dimensional terms allowing insight with a general method

Owen Ren,[a,b] Mohamed Ali Boussaidi,[a,c] Dmitry Voytsekhovsky,[b] Manabu Ihara[d] and Sergei Manzhos[d,1]

[a] Centre Énergie Matériaux Télécommunications, Institut National de la Recherche Scientifique, 1650 boulevard Lionel-Boulet, Varennes QC J3X1S2 Canada.

[b] PureFacts Inc., 48 Yonge St Suite 400, Toronto, ON M5E 1G6 Canada.

[c] Ecole Nationale d'Ingénieurs de Tunis, Rue Béchir Salem Belkhiria Campus universitaire, BP 37, 1002, Le Bélvédère, Tunis, Tunisia.

[d] School of Materials and Chemical Technology, Tokyo Institute of Technology, Ookayama 2-12-1, Meguro-ku, Tokyo 152-8552 Japan

.

[1] Corresponding author. E-mail: manzhos.s.aa@m.titech.ac.jp

**Brief summary of Gaussian process regression (GPR)**

Given the set of samples $f^j$ of a function $f(\boldsymbol{x})$ at points in space $\boldsymbol{x}^j$, $j = 1, \ldots, M$, the expectation values $f(\boldsymbol{x})$ and variances $\Delta f(\boldsymbol{x})$ of function values at other points in space $\boldsymbol{x}$ are computed as[1]

$$
\begin{aligned}
f(\boldsymbol{x}) &= \boldsymbol{K}^*\boldsymbol{K}^{-1}\boldsymbol{f} \\
\Delta f(\boldsymbol{x}) &= K^{**} - \boldsymbol{K}^*\boldsymbol{K}^{-1}\boldsymbol{K}^{*T}
\end{aligned}
\tag{S1}
$$

where $\boldsymbol{f}$ is a vector of all $f^j$ values, and the matrices $\boldsymbol{K}$ and $\boldsymbol{K^*}$ are computed from pairwise covariances among the data:

$$
\boldsymbol{K} = \begin{pmatrix} k(\boldsymbol{x}^{(1)}, \boldsymbol{x}^{(1)}) + \delta & k(\boldsymbol{x}^{(1)}, \boldsymbol{x}^{(2)}) & \cdots & k(\boldsymbol{x}^{(1)}, \boldsymbol{x}^{(M)}) \\ k(\boldsymbol{x}^{(2)}, \boldsymbol{x}^{(1)}) & k(\boldsymbol{x}^{(2)}, \boldsymbol{x}^{(2)}) + \delta & & k(\boldsymbol{x}^{(2)}, \boldsymbol{x}^{(M)}) \\ \vdots & & \ddots & \vdots \\ k(\boldsymbol{x}^{(M)}, \boldsymbol{x}^{(1)}) & k(\boldsymbol{x}^{(M)}, \boldsymbol{x}^{(2)}) & \cdots & k(\boldsymbol{x}^{(M)}, \boldsymbol{x}^{(M)}) + \delta \end{pmatrix}
\tag{S2}
$$

$$
\boldsymbol{K}^* = \left( k(\boldsymbol{x}, \boldsymbol{x}^{(\boldsymbol{1})}) \quad k(\boldsymbol{x}, \boldsymbol{x}^{(\boldsymbol{2})}) \quad \ldots \quad k(\boldsymbol{x}, \boldsymbol{x}^{(\boldsymbol{M})}) \right), \quad K^{**} = k(\boldsymbol{x}, \boldsymbol{x})
$$

The covariance function $k(\boldsymbol{x}_1, \boldsymbol{x}_2)$ is the kernel of GPR. The optional $\delta$ on the diagonal has the meaning of the magnitude of Gaussian noise and is a regularization parameter; it helps generalization.

The covariance function is usually chosen as one of the Matern family of functions[2] given by

$$
k(\boldsymbol{x}, \boldsymbol{x}') = A \frac{2^{1-\nu}}{\Gamma(\nu)} \left( \sqrt{2\nu} \frac{|\boldsymbol{x} - \boldsymbol{x}'|}{l} \right)^{\nu} K_{\nu} \left( \sqrt{2\nu} \frac{|\boldsymbol{x} - \boldsymbol{x}'|}{l} \right)
\tag{S3}
$$

At different values of ν this function becomes a Gaussian-like ($\nu \to \infty$), a simple exponential ($\nu = 1/2$) and various other widely used functions (such as Matern3/2 and Matern5/2 for ν = 3/2 and 5/2, respectively). It is typically preset by the choice of the kernel, and the length scale $l$ and prefactor $A$ are hyperparameters that can be optimized.

**Figures**

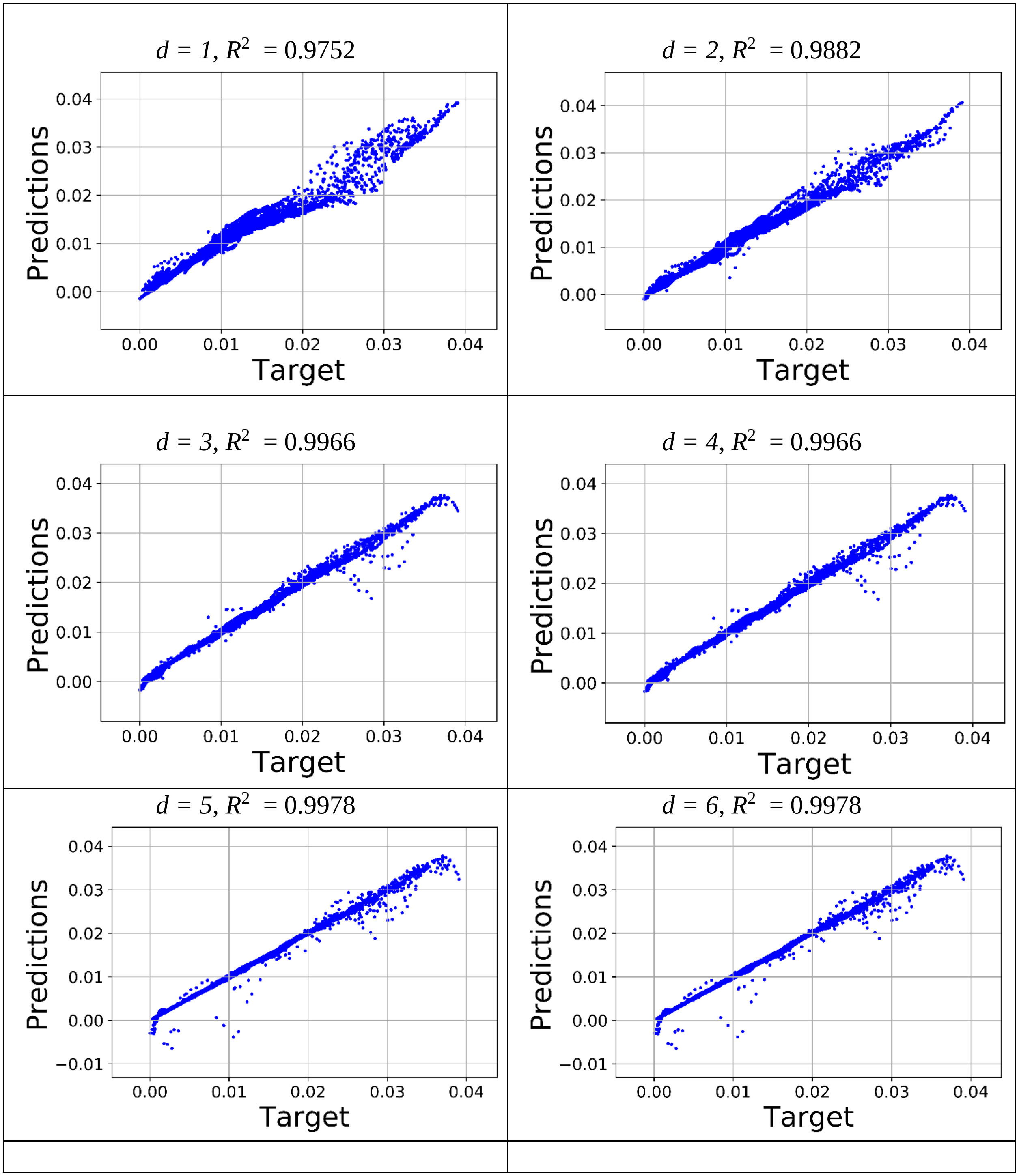
d = 1, R2 = 0.9752
d = 2, R2 = 0.9882
d = 3, R2 = 0.9966
d = 4, R2 = 0.9966
d = 5, R2 = 0.9978
d = 6, R2 = 0.9978
Predictions
Target
0.00 0.01 0.02 0.03 0.04
−0.01

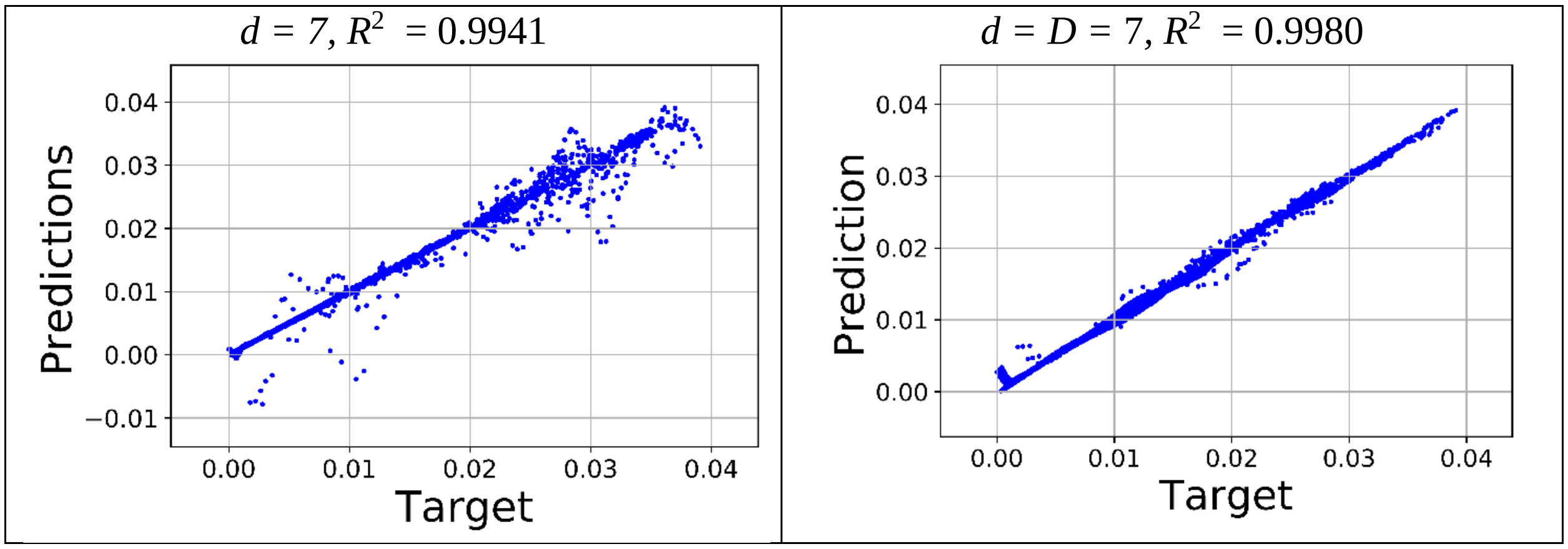

Figure S1. Correlation plots for of HDMR-GPR expansions of orders up to $d$, as well as for a full-dimensional GPR ($d = D = 7$) when fitting kinetic energy densities using 2,000 training points. The plots and Pearson $R^2$ correlation coefficients are for all available data, about 586,000 points.

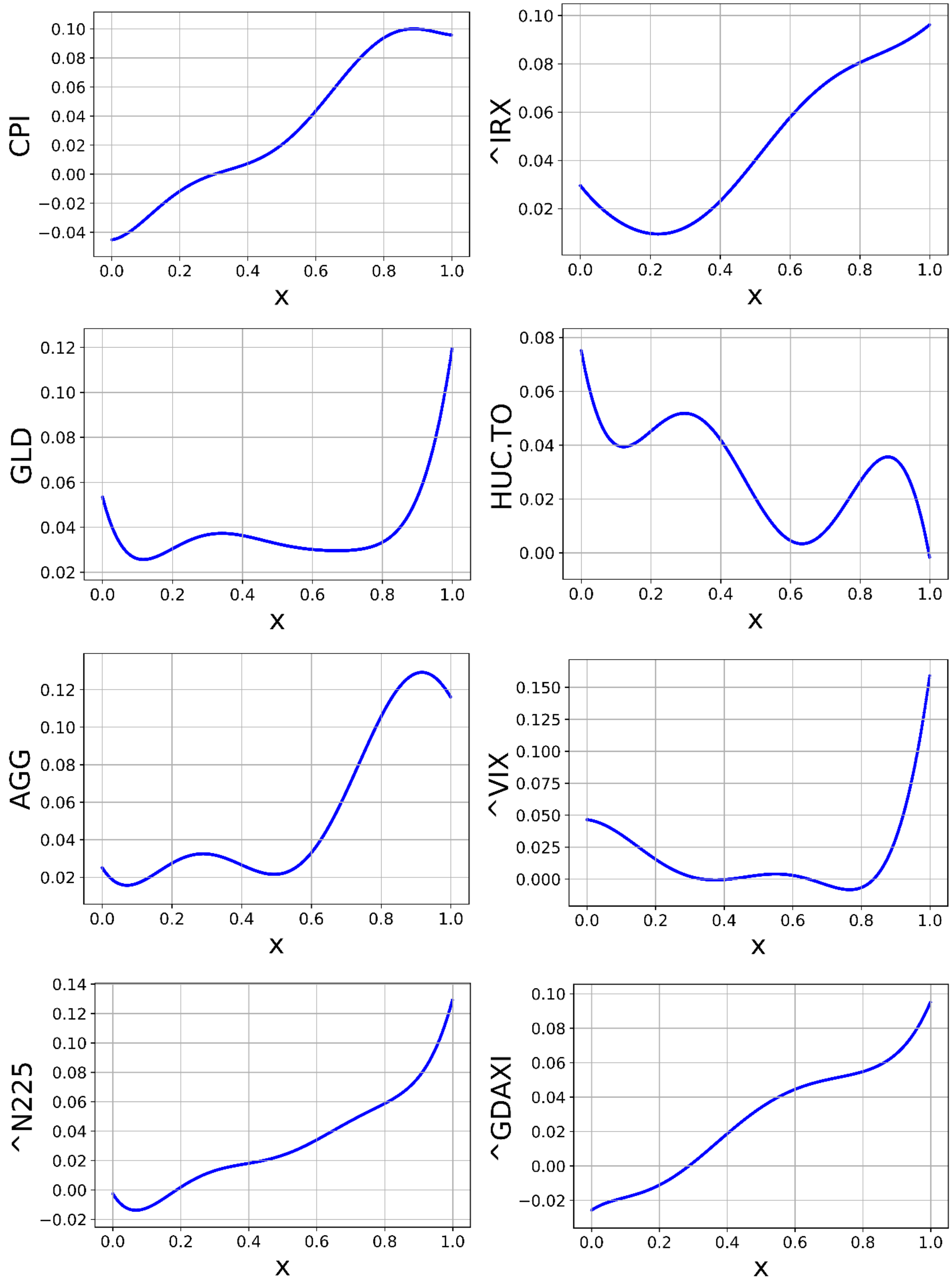

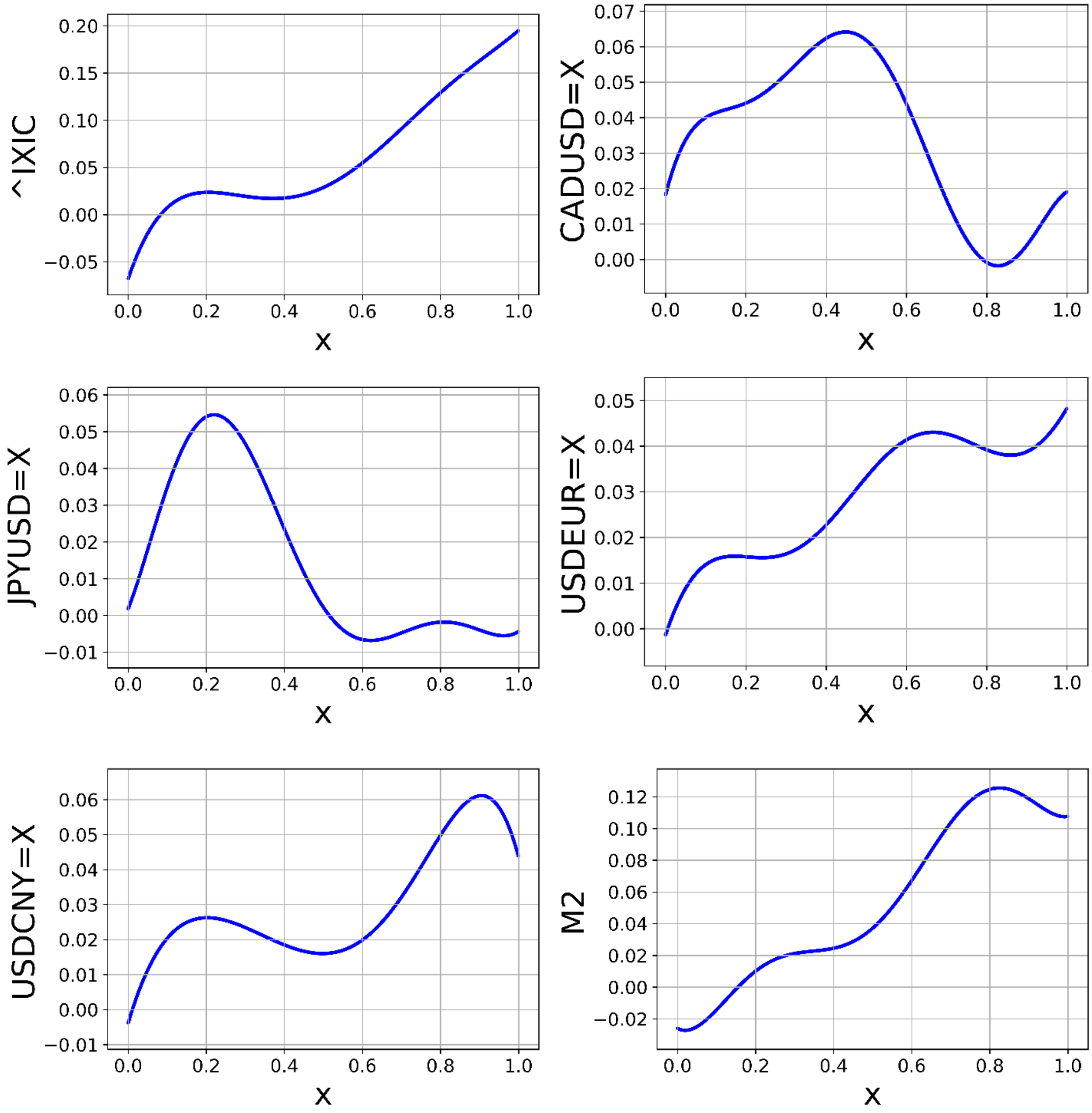

Figure S2. Component functions of the 1$d$-HDMR-GPR model for the example of section 3.4.

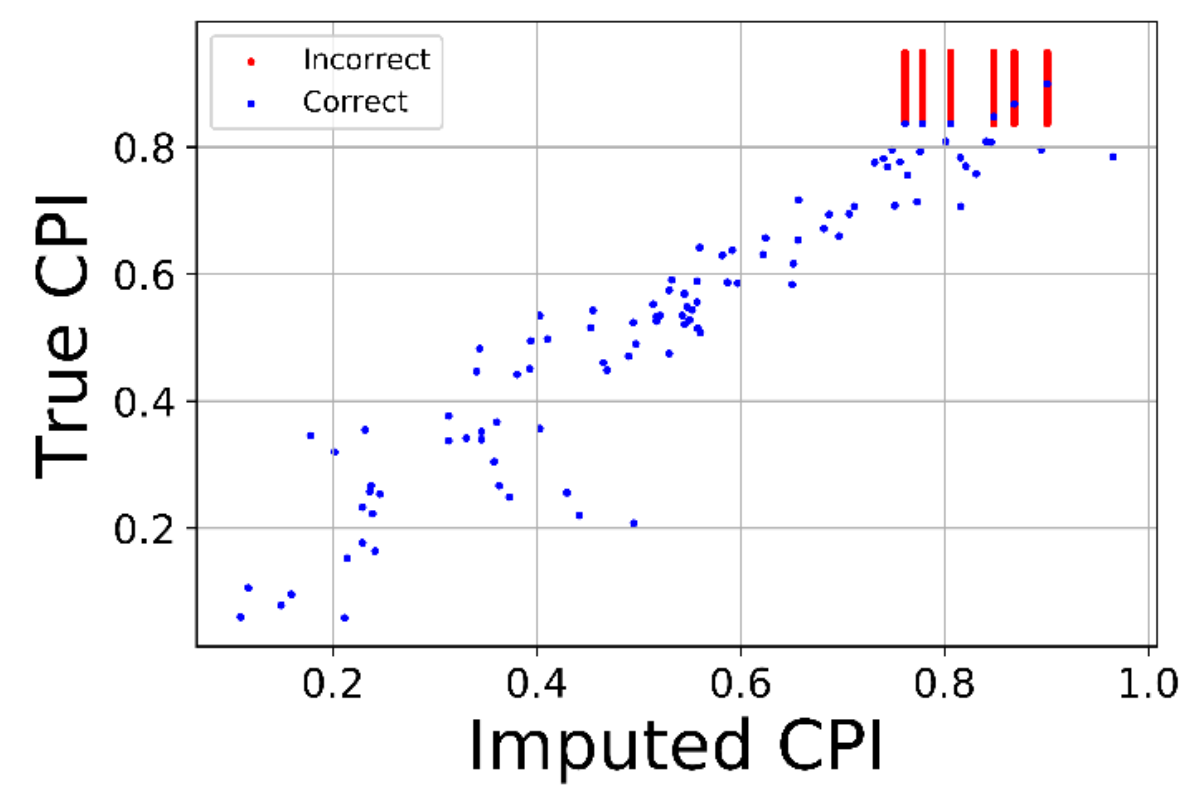
Incorrect
Correct
True CPI
Imputed CPI

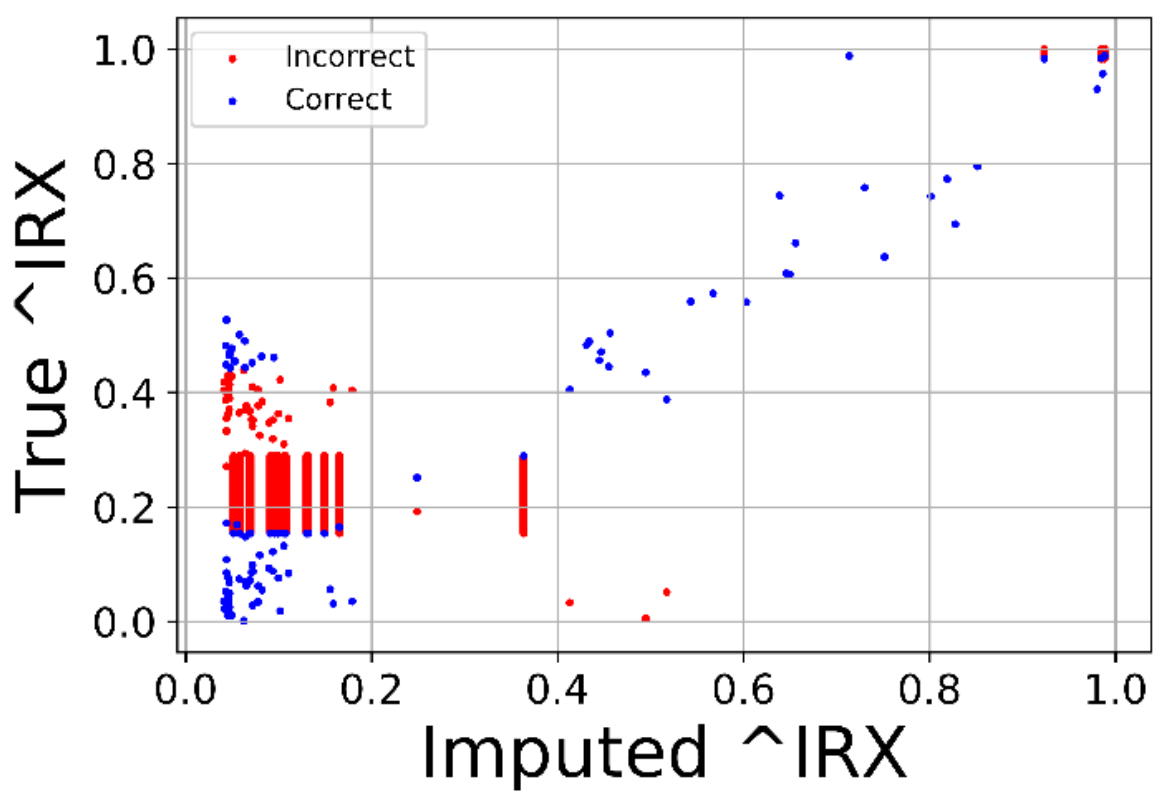
Incorrect
Correct
True ^IRX
Imputed ^IRX

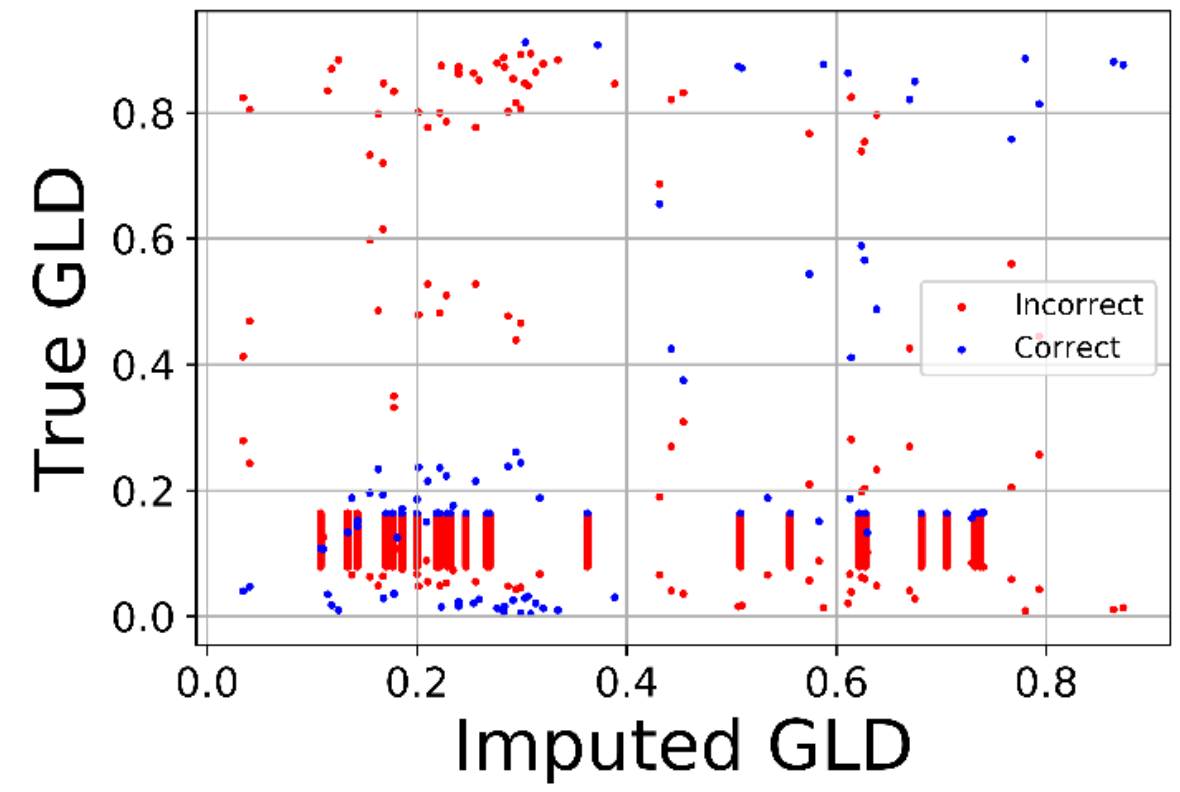
Incorrect
Correct
True GLD
Imputed GLD

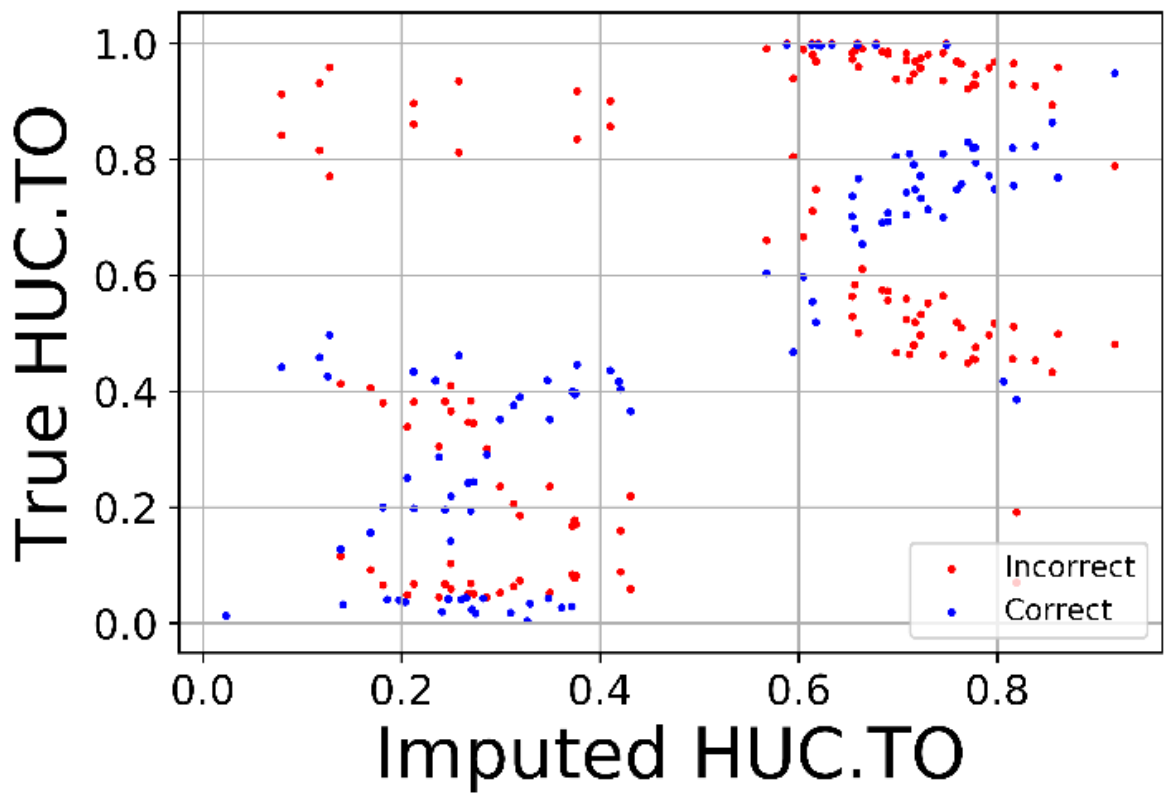
Incorrect
Correct
True HUC.TO
Imputed HUC.TO

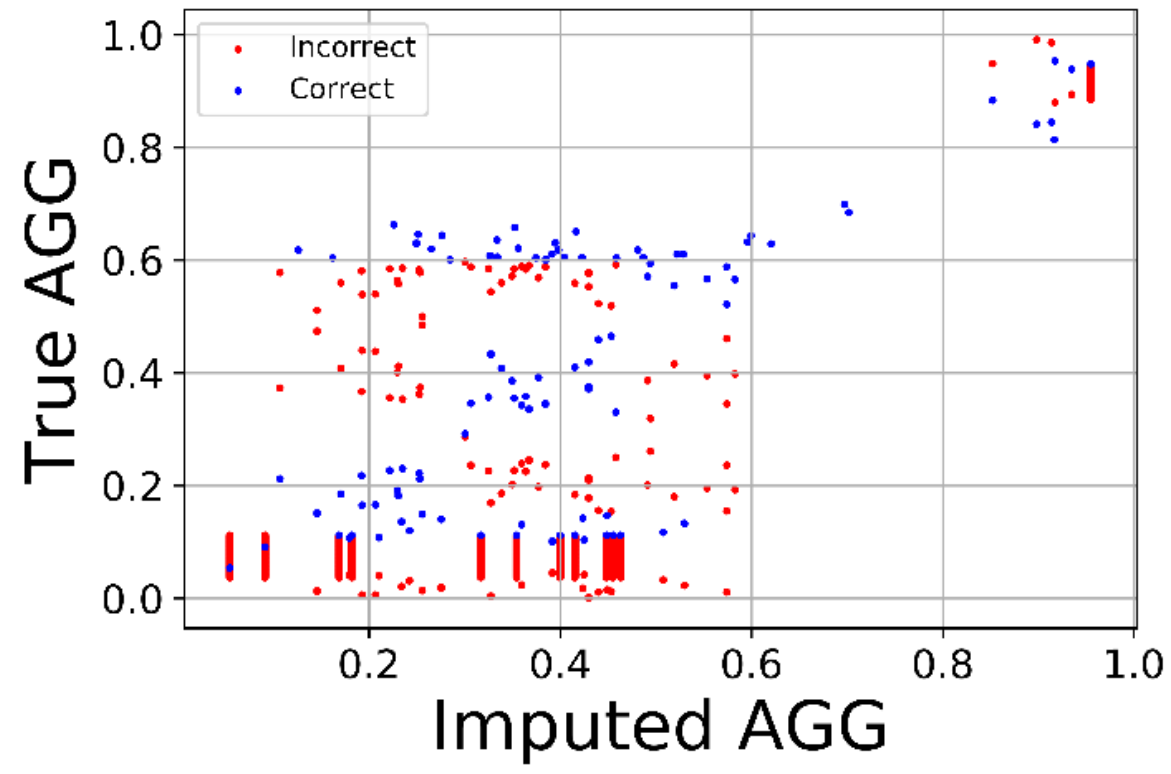
Incorrect
Correct
True AGG
Imputed AGG

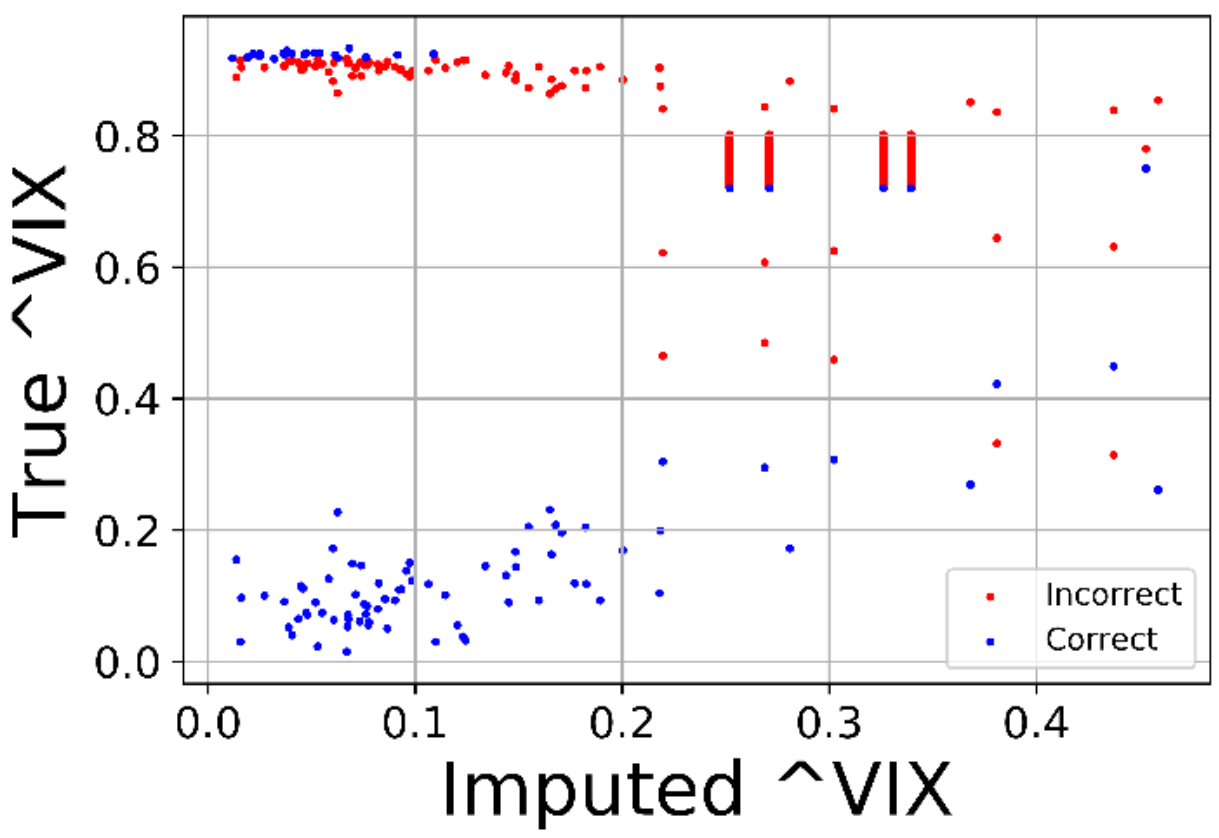
Incorrect
Correct
True ^VIX
Imputed ^VIX

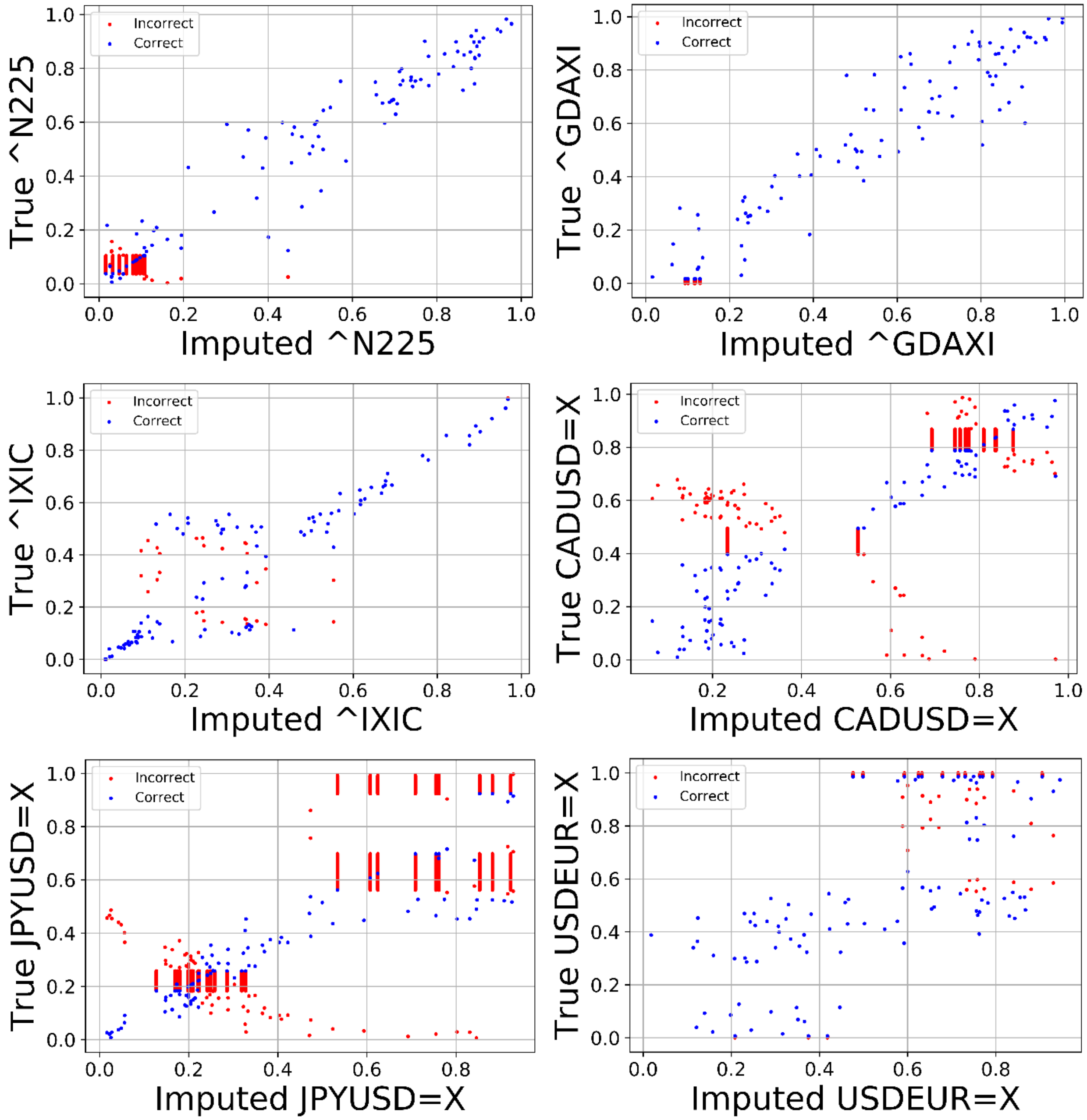

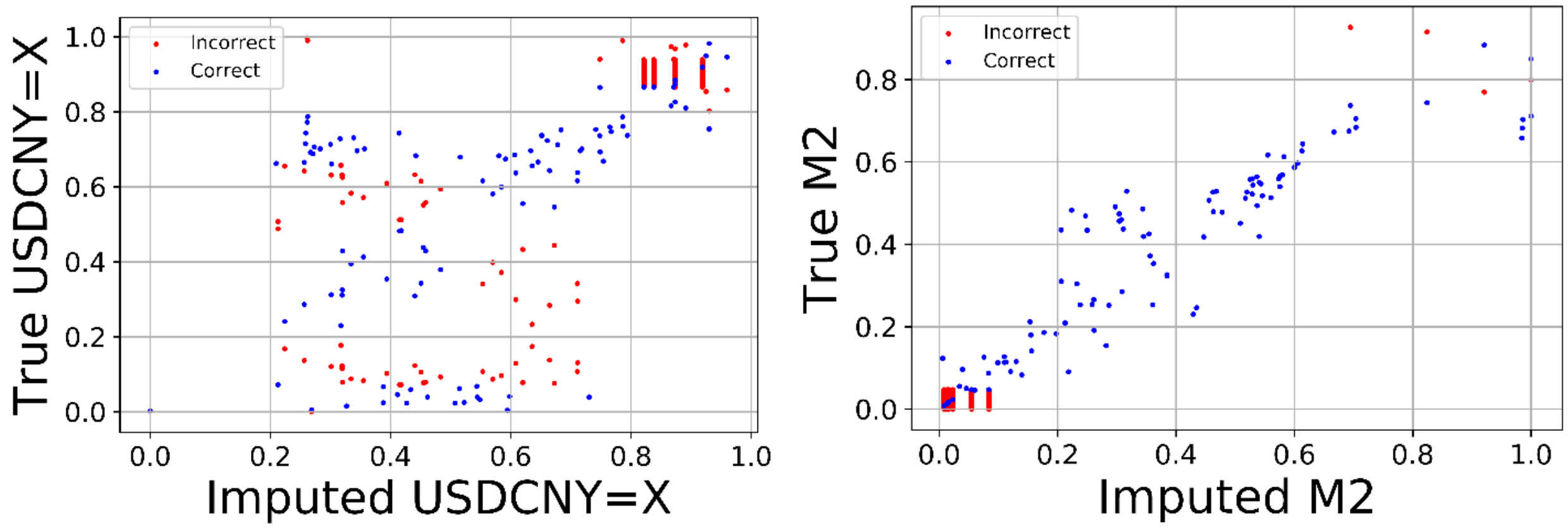

Figure S3. True vs. imputed values of variables for the example of section 3.4.